\documentclass[lettersize,journal]{IEEEtran}

\usepackage{circledsteps}
\usepackage{hyperref}%V
\hypersetup{
 colorlinks,
 citecolor=gray,
 linkcolor=purple,
}

\usepackage{url}%V
\usepackage{mathtools}%V
\usepackage{amsmath}%V
\usepackage{amsthm}%V
\usepackage{xspace}%V
\usepackage{hhline}%V
\usepackage{balance}%V
\usepackage{algorithm}%V
\usepackage{algpseudocode}%V
\usepackage{comment}%V
\usepackage{thm-restate}%V
\usepackage{dsfont}%V
\usepackage{float}%V
\usepackage{subcaption}
\usepackage{slashbox}
\usepackage{amssymb}
\newcommand{\model}{\textsc{DTG-GPT}\xspace}
\newcommand{\dd}{d_{model}}
\newcommand{\fil}{f}
\newcommand{\token}{\textsc{Tok}}

\newcommand{\seg}{s}
\newcommand{\traceN}{tr}

\newcommand{\matSize}{0.121}

\newcommand{\para}[1]{\vspace{2pt} \noindent \textbf{#1}\xspace}

\newcommand{\temp}{t}

\newcommand{\card}[1]{\lvert #1 \rvert}

\makeatletter

\patchcmd{\maketitle}
	{\@maketitle}
	{\@maketitle\vspace{-0em}}% change the value as needed
	{}
	{}
\makeatother

\usepackage{subfiles} % Best loaded last in the preamble
 
\begin{document}

\title{ Harnessing Generative Pre-Trained Transformer for Datacenter Packet Trace Generation}

\author{
    \IEEEauthorblockN{Chen Griner }\\
    \IEEEauthorblockA{{Independent researcher, formerly with Ben-Gurion University of the Negev \\ grinerchen@gmail.com  
     } 
}}
 
 \maketitle 
\begin{abstract}
Today, the rapid growth of applications reliant on datacenters calls for new advancements to meet the increasing traffic and computational demands. Traffic traces from datacenters are essential for further development and optimization of future datacenters. However, traces are rarely released to the public. Researchers often use simplified mathematical models that lack the depth needed to recreate intricate traffic patterns and, thus, miss optimization opportunities found in realistic traffic. 
In this preliminary work, we introduce \model, a packet-level Datacenter Traffic Generator (DTG), based on the generative pre-trained transformer (GPT) architecture used by many state-of-the-art large language models.    
We train our model on a small set of available traffic traces from different domains and offer a simple methodology to evaluate the fidelity of the generated traces to their original counterparts.
We show that \model can synthesize novel traces that mimic the spatiotemporal patterns found in real traffic traces. 
We further demonstrate that \model can generate traces for networks of different scales while maintaining fidelity.  
Our findings indicate the potential that, in the future, similar models to \model will allow datacenter operators to release traffic information to the research community via trained GPT models.  
\end{abstract}
% \begin{IEEEkeywords}
% Traffic modeling and synthesis; datacenters; 
% \end{IEEEkeywords}
\section{Introduction}\label{sec:intro}
 
Datacenter networks (DCNs) constitute a critical component of the infrastructure supporting the contemporary digital world. Large-scale web applications, cloud computing platforms, and, in particular, the rapidly growing deployment of generative AI models all require increasing efficiency and computing power. Thus, in the coming years, an explosive increase in the demand for datacenters is expected \cite{dataCenterGrowth2023URL,Open_Compute_Project2022}. %
Novel technologies and developments will likely be needed for the next evolution in DCNs. %Indeed, the computational costs of the newest models are already prohibitive for a single task \cite{o3modelcost}.
While there have been many recent developments from researchers in datacenter networking \cite{hall2021survey}, an important challenge to progress is the lack of realistic datacenter network traces that would be used in the development of these technologies \cite{avin2019toward,adeleke2022networksurvey,kong2024high,foerster2019survey}.
Traditional DCNs are oblivious to traffic and have a static topology; their design often assumes a uniform distribution of traffic~\cite{avin2019toward,complexity2020}. Some novel technologies for DCNs take a different approach \cite{adeleke2022networksurvey,projector,dan}. These networks, called demand-aware networks, can change their topology and adjust to a changing traffic pattern, thus increasing efficiency.
The lack of realistic traffic traces hinders progress for novel demand-aware networks, since network designers remain oblivious to the actual traffic patterns in modern datacenters. 
However, these tracks are usually available only to network operators and not to researchers \cite{adeleke2022networksurvey}.  
The lack of traces, except for a few examples, is often caused by privacy concerns. Indeed, even simple anonymization of the trace may not be enough to release the trace \cite{sicker2007legal}. 
Lastly, even if traces were available, replaying them in a simulation would likely only benefit simulated networks of an identical topology \cite{adeleke2022networksurvey}. 

The networking community has devised many synthetic traffic generators to address these issues and others. However, it is not clear to what extent they mimic realistic traffic patterns, nor is there a good framework to decide which traces are a good emulation of realistic network traffic \cite{adeleke2022networksurvey}.
Indeed, it is likely that different types of DCNs running different tasks would each have different traffic patterns \cite{complexity2020}.

Recently, several works have appeared that use different generative AI methods to generate traces. For example, \cite{cheng2019pac, yin2022practical,byun2024generative,ring2019flow} use GANs (based on convolutional recurrent networks, convolutional long short-term memory) to generate packet traces.  % they create a traffic figure from a trace a
A few papers have emerged that use a generative pre-trained transformer (GPT) based architecture for traffic generation. In \cite{meng2023netgpt}, the authors utilize the GPT architecture to generated traces of DNS and cryptocurrency traffic.  In \cite{qu2024trafficgpt}, the authors attempt to reconstruct simple packet traces using a pre-trained GPT-3 model, which they fine-tune to generate packet traces containing ICMP and DNS packets. In \cite{kong2024high}, the authors use a GPT model to generate traces that mimic control-plane traffic of 4G and 5G cellular networks.
However, we are unaware of any work dedicated to generating and evaluating realistic DCN traces using the GPT architecture. 
In this work, we present \model, a novel DCN trace generation model based on the GPT architecture used by large language models such as OpenAI Chat-GPT and others \cite{vaswani2017attentionisallyouneed}. 
In this initial work, we will focus on a small but crucial part of the DCN packet trace: the source and destination IP columns. In this manner, we follow an example set in previous works looking at DCN traces for DCN development \cite{complexity2020, griner2024beyond,bienkowski2021online}.

\begin{figure*}[!h]
  \begin{centering}
  \begin{tabular}{cc}
  \includegraphics[width=.35\linewidth]{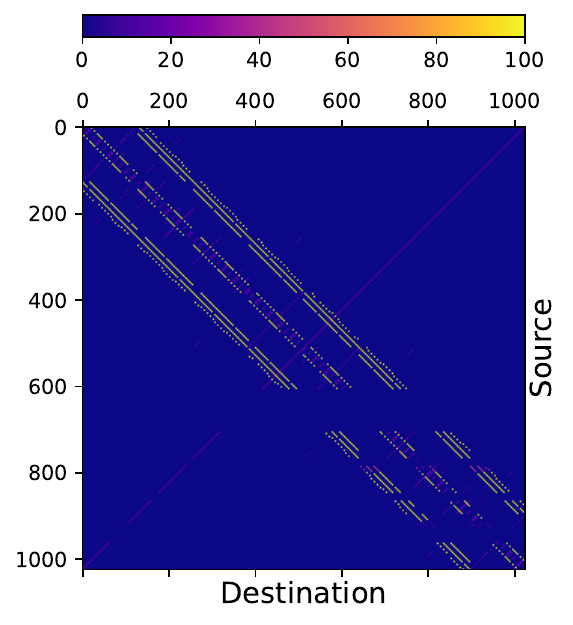}&
 
  \includegraphics[width=.35\linewidth]{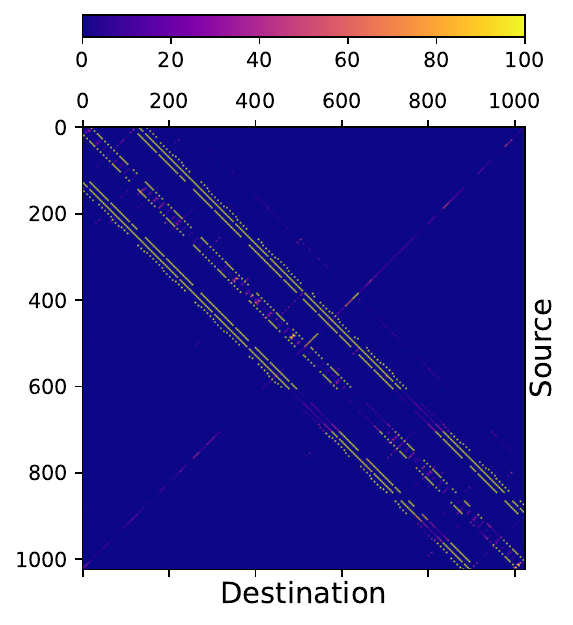} \\
   \small{(a) Original} & \small{(b) Generated}
  \end{tabular}
   \caption {Two traffic matrices of $8M$ requests. Where (a) is a real trace from an HPC cluster (NeckBone) \cite{doe2016characterization} and (b) is a counterpart trace generated with \model. The maximal element was clipped to $100$ requests to enhance contrast.  }
    \label{fig:realandfakeEample}
  \end{centering}
  %\vspace{-.4cm}
\end{figure*}
 
Generating complex, varied, and realistic DCN traces is not a straightforward task. Many variables exist, such as the network size, the datacenter's task, the protocols used, etc. However, as a first step, we would like to mimic existing traces. That is, given a network trace (which stands for a traffic pattern), we would like our model, \model, to generate more traces, which are similar to the original trace in a fundamental way but not identical, essentially making more of the same trace.
To judge the fidelity of the novel traffic traces generated with \model, we would like to compare them to the original. However, comparing two traces is not very straightforward; there could be many ways to measure and compare traffic \cite{adeleke2022networksurvey}.
We begin with a simple approach: we look at \emph{traffic matrices}. These matrices describe the accumulated traffic between each source and destination in the network, as seen in the trace. This allows us to observe the spatial structure of the traffic that arises from the frequency of all source-destination pairs.
We illustrate our intention with the following example.

Figure \ref{fig:realandfakeEample} shows two traffic matrices. Figure \ref{fig:realandfakeEample} (a) shows the total traffic matrix from $8M$ request long trace (denoted as 'Neckbone' in the original dataset), and Figure \ref{fig:realandfakeEample} (b)
shows a synthetic counterpart trace of $8M$ requests generated using our \model model.
As we can see, a clear pattern emerges: three main diagonals go from left to right, and each diagonal is broadly composed of three smaller diagonals. There are also other features, such as an empty gap in the diagonals lacking traffic between sources $600$ and $700$ and a smaller diagonal going from right to left.  The reader may look closer at the two matrices and find many smaller similarities and, crucially, some differences.
We note that part of our dataset is composed of traces that describe HPC traffic (see Section \ref{subsec:data}), and this traffic has a clear structure, which can be seen in the accumulated traffic matrix. 
However, looking at different structures or patterns in the traffic matrix can only give us a partial indication of similarity; it does not consider a trace's temporal properties. Indeed, It is possible to generate a trace with a very similar traffic matrix to Figure \ref{fig:realandfakeEample} (a) by sampling from a probability distribution based on the frequency of all pairs in the traffic matrix.

Throughout this work, we will therefore attempt to show that our \model generates not only traces with similar spatial properties but also temporal ones, and it does so while generating novel sequences based on the original traces.

Our contributions are as follows. 
We present \model, a GPT-based machine learning model, able to generate novel DCN packet traces based on existing traces.
We discuss a different tokenization strategy for traffic traces and offer some initial changes that make a GPT model fit for natural language processing to a model able to generate traffic traces. 
We investigate a methodology that can help estimate the fidelity of the generated traces to existing traces. We then evaluate the model's trace generation abilities for seven traces from different domains and network scale sizes. We demonstrate that traces generated by \model show a high degree of novelty, which may help solve some concerns regarding the release of traffic trace information from network operators.

The rest of this paper is organized as follows: We begin by presenting our trace and network models and some preliminaries in Section \ref{sec:perlim}, and we shortly present the architecture of our ML model in Section  \ref{sec:Architecture} (as well as our transition scheme). In the evaluation in Section \ref{sec:eval}, we describe the training process. We show results for several generated traces. We conclude in Section \ref{sec:FutureWork} and present ideas for future work.  

%============================================================
\section{Preliminaries}\label{sec:perlim}
This section briefly discusses our traces and network model and provides background on trace complexity, which we use in the technical sections. 
\subsection{Traffic \& Trace models}\label{subsec:Traffictracemodel}
 
\para{Network}
Typically, a network is defined via its topology as a set of vertices and nodes that form a graph. However, in the context of this paper, we are agnostic to the network's structure in terms of its particular set of edges. We are only required to know its set of nodes. The scale of a network is the number of nodes that comprise the network, denoted by $n$. The ID of each node is in the range $[0, n-1 ]$. We will also assume that any pair of nodes in the network may communicate.   

\para{A Network Trace}
A trace $\sigma$  is a sequence of requests $\sigma_i=\{\sigma_1,\sigma_2,\dots,\sigma_{\card{\sigma}}\}$, of length $\card{\sigma}$, where each request represent a single packet of traffic. In this work, each request in the trace contains the source and destination node IDs $\sigma_i = ( s_i,d_i) $,  $s_i$, and $d_i$, respectively. This follows some previous works which used datacenter traffic traces \cite{complexity2020, griner2024beyond,bienkowski2021online}.

\para{Traffic Matrix}
We denote a traffic matrix as $M$. This is an $n\times n$ matrix, representing the demand volume between each source and destination pair, that is, the number of requests. More formally, for every source and destination $s,d \in [0,n-1]$, the value of every cell $M[s,d]$ in the matrix $M$ represents the number of occurrences of the request $(s_i,d_i)$ in the trace $\sigma$.
% \para{Demand graph}
% The demand graph $G(M)=\{V,E\}$ is a graph generated directly from the demand matrix by assuming it is the adjacency matrix. This graph has $n=\card{V}$ nodes, and $\card{E}$ edges. 
% an edge between two nodes $s$ and $d$ exists if the entry $M[s,d]$ of the demand matrix is greater than zero $M[s,d]>0$.
%%%%%%%%%%%%%%%%%%%%%%%%%%%%%%%%%%%%%%%%%%%%%%%%%%%%%%%%%%%%%%
%%%%%%%%%%%%%%%%%%%%%%%%%%%%%%%%%%%%%%%%%%%%%%%%%%%%%%%%%%%%%%
%%%%%%%%%%%%%%%%%%%%%%%%%%%%%%%%%%%%%%%%%%%%%%%%%%%%%%%%%%%%%%
%%%%%%%%%%%%%%%%%%%%%%%%%%%%%%%%%%%%%%%%%%%%%%%%%%%%%%%%%%%%%%
%%%%%%%%%%%%%%%%%%%%%%%%%%%%%%%%%%%%%%%%%%%%%%%%%%%%%%%%%%%%%%
%%%%%%%%%%%%%%%%%%%%%%%%%%%%%%%%%%%%%%%%%%%%%%%%%%%%%%%%%%%%%%
\subsection{Trace Complexity}\label{subsec:perlim:comp}
Part of the toolbox we will use to judge the fidelity of our generated traces is \emph{trace complexity} \cite{complexity2020}. This approach can identify and quantify the types of structure featured by packet traces in communication networks, particularly data centers. \emph{Trace complexity}, approximates information theory's measure of \emph{entropy rate} ~\cite{shannon1948mathematical} for traces\footnote{Since the term entropy is defined for random variables, as opposed to a sequence of individual communication requests in a packet trace, the term complexity is used similarly to earlier work on random sequences, such as in this work by Lempel and Ziv~\cite{lempel1976complexity}, where the complexity of a random sequence is linked to its similarity to a random sequence. }. By estimating the entropy rate, it can quantify the amount of \emph{temporal} and \emph{non-temporal} structure in traffic traces.

Briefly, trace complexity gives an information-theoretic measure on the \emph{structure}, or conversely, randomness, found in a traffic trace.  
Intuitively, a packet trace with \emph{low entropy} has \emph{low complexity}: it contains little information, and the sequence behavior is more predictable \cite{feder1992universal}; hence, it has \emph{high structure}. Conversely, a more random trace will have \emph{high complexity} and  \emph{low structure}. Importantly, it is possible to measure different aspects of a trace's complexity by randomizing different aspects of the trace while keeping the rest intact. 

For the purpose of this paper, we explore two types of trace complexity: temporal complexity and non-temporal complexity. Concisely, temporal complexity measures the randomness represented by the ordering of the packets (e.g., bursts, repetitions, cycles of similar elements, etc.). Non-temporal complexity estimates the randomness represented by the frequency requests or nodes.

For a full discussion of trace complexity, we refer the reader to the original paper~\cite{complexity2020}

We believe that trace complexity offers a simple tool to estimate the fidelity of our generated trace. That is, if the generated traces are similar to their original counterparts, they should have similar values in terms of both temporal and non-temporal complexities. In particular, temporal complexity measures an aspect of a sequence that is otherwise difficult to quantify. Furthermore, since trace complexity is a normalized measure, it can compare traces of different lengths and node counts.

%%%%%%%%%%%%%%%%%%%%%%%%%%%%%%%%%%%%%%%%%%%%%%%%%%%%%%%%%%%%%%
%%%%%%%%%%%%%%%%%%%%%%%%%%%%%%%%%%%%%%%%%%%%%%%%%%%%%%%%%%%%%%
%%%%%%%%%%%%%%%%%%%%%%%%%%%%%%%%%%%%%%%%%%%%%%%%%%%%%%%%%%%%%%
%%%%%%%%%%%%%%%%%%%%%%%%%%%%%%%%%%%%%%%%%%%%%%%%%%%%%%%%%%%%%%
%%%%%%%%%%%%%%%%%%%%%%%%%%%%%%%%%%%%%%%%%%%%%%%%%%%%%%%%%%%%%%
%%%%%%%%%%%%%%%%%%%%%%%%%%%%%%%%%%%%%%%%%%%%%%%%%%%%%%%%%%%%%%
\section{\model Architecture}\label{sec:Architecture}
\begin{figure}[h]
  \begin{centering}
  
  \includegraphics[width=0.47\textwidth]{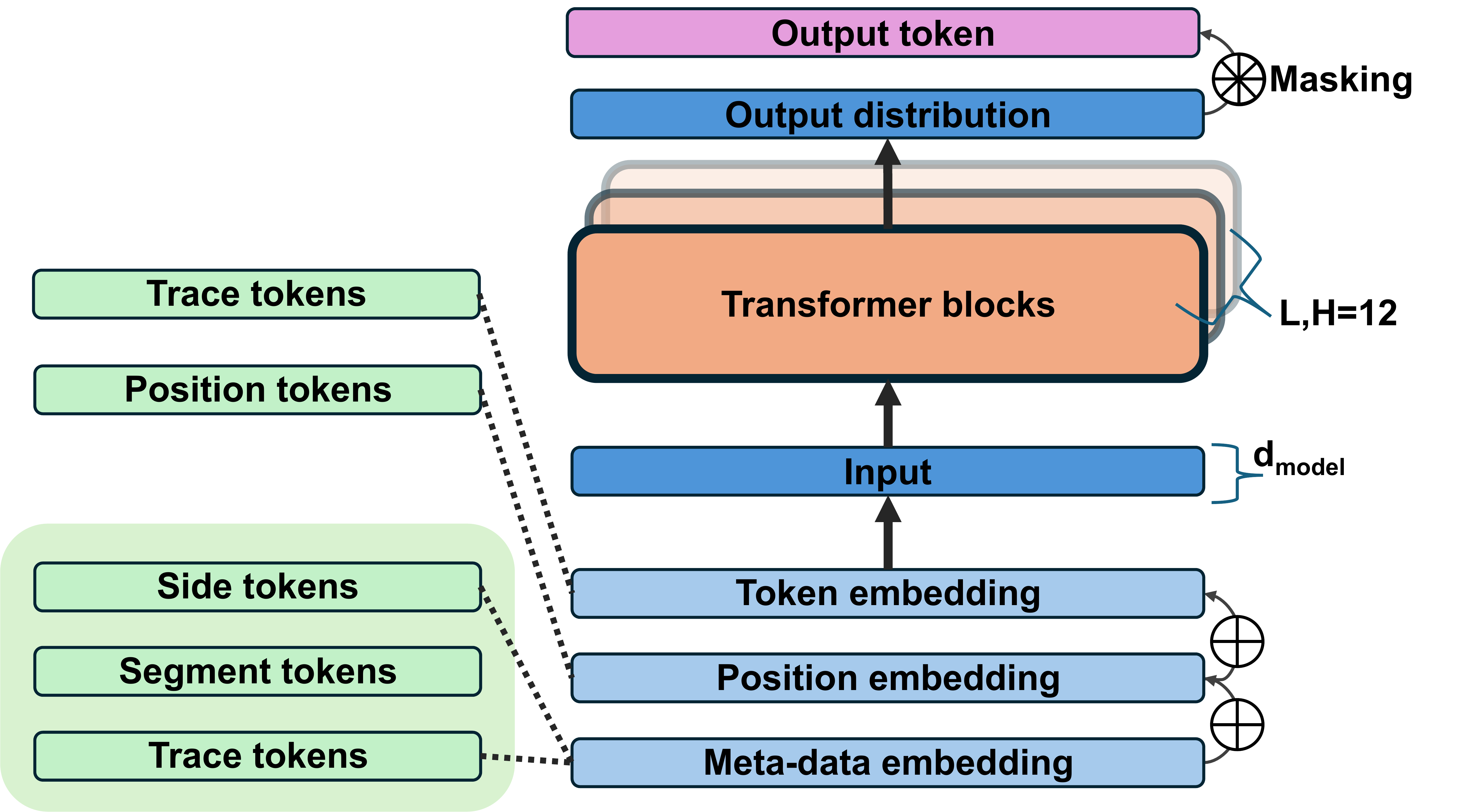}

    \caption{Simplified scheme for our \model model .   
    }
    \label{fig:modelarchi}
  \end{centering}
\end{figure}
In this section, we will discuss the architecture of our model.
Figure \ref{fig:modelarchi} shows a simplified schematic of \model. 
Briefly, our model is based on the GPT architecture, with the main changes being for the input embeddings. Here, the input to the transformer block is the sum of three embedding layers. For trace generation, the output distribution is used, after possibly masking it, to create a single output token.

The core part of our model is the transformer block \cite{vaswani2017attentionisallyouneed}. More specifically, for this part, our design and code are based on OpenAI's GPT-2 large language model \cite{GPT22019language, radford2018improvingGPT}. 
For the transformer block, there are two parameters: $L$ represents the number of linear layers, $H$ denotes the number of self-attention heads, and the size of the embedding dimension is denoted as $\dd$. 
We report on the exact parameters used for the transformer block in our evaluation in Section \ref{subsec:trainsetup}.

Our model uses a different set of embedding layers. The original GPT-2 model has two embedding layers: the token embedding, which is used to embed tokens from the original data (e.g., text) and the location embedding, which represents the position of each token, as described in previous works \cite{vaswani2017attentionisallyouneed,vaswani2017attentionisallyouneed}\footnote{ The BERT model also adds a segment embedding layer \cite{devlin2018bert}.}.
In our model, similarly to GPT-2, each token in the token vector represents either a source or a destination ID from the trace (we further elaborate on this in Section \ref{subsec:tokenz}). %is embedded into the token embedding layer. 
The position vector represents the position of each token in the token vector. 
Each vector is embedded into its respective embedding layer. 
Unlike GPT-2, our \model is modified to contain another embedding layer, which we denote as the meta-data embedding layer, which makes it better suited to the task of network trace generation.  

The addition of this layer is motivated by the following observations: the order of the fields of a packet header in a network trace is deterministic. For example, a source ID is \textit{always} followed by a destination ID, etc. However, the model, which is fit to generate natural language, will not naturally be aware of this fact. 
Furthermore, we may want to generate a trace of a specific type or a specific part of a trace, which may not be naturally possible. 
Thus, adding the meta-data embedding allows us more control and better training of the model.

The meta-data embedding layer encodes additional information regarding each token. This allows us to add information about each token to the model without explicitly using a token; as we will see, this information can then be controlled when using the model to generate novel traces from a specific type. 
The meta-data vector is constructed using three different vectors, which are then transformed into a single meta-data vector. These are the segment vector, the field vector, and the trace ID vector.
 
The \emph{field vector} represents the packet header field to which the corresponding token belongs, helping the model deal with the deterministic nature of the packet header. Each token $\token^f_i$ is an integer from the range $\token^f_i \in [0, \fil-1]$, where $\fil$ is the number of fields in the packet header. Since, in this paper, we only consider the source and destination fields, these are represented with a $0$ and a $1$, respectively, in the vector and $\fil=2$. Since a destination ID can only follow a source ID and vice versa, this vector is always of the form $\{\dots 0,1,0,1,0\dots\}$ for any input. We believe that this vector helps the model learn source-destination dependencies.
 
The segment and trace ID vectors help the model learn to which part (or segment) of the trace and to which trace each token belongs.
Their addition is motivated by the following observation.
The \textit{context length} of large language models determines the maximal number of tokens that the model can process at a single 'pass'; these tokens are its context window. Since the model's predictions are based only on the current tokens in the window, it is typically assumed that a larger context length leads to more accurate predictions; essentially, since the model has more 'context' to consider, we, therefore, would this value to be as large as possible.  
However, increasing the context length has a significant performance penalty. 

Often, natural language generation tasks are shorter, only a few hundred tokens at a time. Furthermore, the exact context (e.g., the subject or domain of the text) may be inferred from the initial prompt given to the model. Indeed, chatbots based on LLMs are sometimes given hidden prompts to generate text of a certain form or subjects \cite{GPT3brown2020language}. This may not always be possible when generating traces. We may not always have parts of a trace, and even then, a trace may have parts that are identical in different parts or traces that have identical parts, making it hard for the model to learn the difference. 

We may still want the model to understand the context regardless of the generated traces's length and the limited context window. 
This work attempts a workaround by adding the segment and trace ID vectors. 

The segment vector divides a trace into $\seg$ segments of equal size (possibly clipping the end to fit). This allows the model to attribute sequences to different trace parts during training. We conjecture this could increase model accuracy when trying to recreate temporal patterns. 
In turn, for generation, the user can create traces that mimic a specific period in the trace. We note that the segments could, for example, denote different periods of the day, hours, minutes, days, etc. With each period denoted by a specific segment.
Each token from the segment vector, $\token^s_i$ is from the range $\token^s_i \in [0, \seg-1]$, where $\seg$ is the number of segments for each trace. During training, we generate a vector with sequential tokens from the range. For a given trace with length $n$ each segment token $i$, appears $\lfloor\frac{n}{\seg}\rfloor$ times sequentially, where $\lfloor\frac{n}{\seg}\rfloor$ is the \emph{segment length}.

Finally, the trace ID vector allows the model to identify the type of trace a token belongs to. 
Each trace ID token $\token^t_i$ is from the range $\token^t_i\in [0, \traceN -1]$ where $\traceN$ is the number of traces used during training. 

Each token $\token^m_i$ in the meta-data vector would be from the range $\token^m_i\in[0,\fil\cdot\seg\cdot\traceN-1]$. This token is created from the field, segment, and trace ID vectors using the following formula.

\begin{align}
    \token^m_i=\token^s_i+\token^f_i\cdot \seg +\token^t_i\cdot \fil \cdot \seg 
\end{align}
This creates a unique set of tokens for each trace, and the resulting meta-data vector is embedded into the meta-data embedding later. The result is added to the token and position embeddings and fed to the model during training. During trace generation, the meta-data vector needs to be constructed according to the needs of the trace generation task.

%%%%%%%%%%%%%%%%%%%%%%%%%%%%%%%%%%%%%%%%%%%%%%%%%%%%%%%%%%%%%%
%%%%%%%%%%%%%%%%%%%%%%%%%%%%%%%%%%%%%%%%%%%%%%%%%%%%%%%%%%%%%%
%%%%%%%%%%%%%%%%%%%%%%%%%%%%%%%%%%%%%%%%%%%%%%%%%%%%%%%%%%%%%%
%%%%%%%%%%%%%%%%%%%%%%%%%%%%%%%%%%%%%%%%%%%%%%%%%%%%%%%%%%%%%%
\section{ Training and Generation Setup}\label{sec:Train}
%%%%%%%%%%%%%%%%%%%%%%%%%%%%%%%%%%%%%%%%%%%%%%%%%%%%%%%%%%%%%%
%%%%%%%%%%%%%%%%%%%%%%%%%%%%%%%%%%%%%%%%%%%%%%%%%%%%%%%%%%%%%%
In this section, we discuss the different aspects of the training and the trace generation setup for our model. We begin by reviewing the dataset.
\subsection{Data}\label{subsec:data}

\begin{table}[t]
\caption{Traces used in the research.}
%\vspace{-0.3cm}
\scriptsize
    \centering
    \begin{tabular}{|l|c|c|c|}
\hline
Type                                                      & \# nodes & \# of entries & \# generated \\ \hline
Facebook  DB \cite{facebook}                        & 324      & 20M  & 2M            \\ \hline
Facebook   WEB                                       & 158     & 20M   & 2M            \\ \hline
Facebook   Hadoop                                    & 371     & 20M   & 2M            \\ \hline
HPC (CNS) \cite{doe2016characterization}                  & 1024    & 1.1M   & 1.1M           \\ \hline
HPC (MultiGrid)                                           & 1024    & 17.9M & 8M            \\ \hline
HPC (MOCFE)                                               & 1024    & 2.7M  & 2.7M          \\ \hline
HPC (Nekbone)                                             & 1024    & 20M  & 8M            \\ \hline
\end{tabular}%
    \label{tab:traces_data}
\end{table}
We use two different trace sets. The first is from a Meta (at the time, Facebook ) datacenter, and the second is an HPC cluster. We will denote these two subsets as Facebook and HPC. 

The HPC trace set is composed out of \emph{Four} traces of exascale applications i.e., computing applications that require at least $10^18$ double precision operations per second) in high-performance computing (HPC) clusters~\cite{doe2016characterization}: CNS, MultiGrid, MOCFE and NeckBone. These represent several different computational kernels of different applications and show the communication pattern between $1024$ CPUs. 
The Facebook dataset consists of three traces collected from three datacenter clusters, each corresponding to a distinct application type: Hadoop (HAD), a Hadoop cluster; web (WEB), comprising servers that handle web traffic; and databases (DB), which include MySQL servers responsible for storing user data and processing SQL queries.
For the purpose of this research, all traces were clipped to be at most $20M$ entries long.

The Facebook traces were also filtered.  
The original Facebook traces set were longer and more verbose, up to $300k$ packets and several different fields, as described in previous works \cite{facebook,complexity2020}. Only the source and destination \textit{rack} columns were extracted and tested for this work. Furthermore, the original trace had an uneven set of source and destination rack ID entries (as the trace was captured only on a single part of the DC). The traces were filtered so that only nodes in the source column could appear in the destination column to create a trace with the same set of source and destination nodes. 

Table \ref{tab:traces_data} summarizes the final lengths and node ID numbers for each trace used in this paper; the lengths are the ones used for training (we discuss the \textit{generated} column in a later section).
The original trace files were downloaded from  \cite{trace-collection}.

\subsection{Fine-Tuning}
In this section, we briefly discuss our approach to fine-tuning.

The training process of LLMs using the GPT architecture typically involves at least two phases. In the first phase, called pre-training, the model is exposed to vast amounts of textual data. This is usually the most cost-intensive part of training. In the second phase, often called the fine-tuning phase, the model can be adjusted or trained on a smaller, task-specific data set to increase its performance for a specific task~\cite{devlin2018bert,GPT22019language,GPT3brown2020language}.
We can draw an analogy between our model, \model, and typical LLMs, where traces represent textual data. Since this paper deals with a novel application of the GPT architecture, in this analogy, the main focus of our effort will be the pre-training phase.
 
We take a simple approach for fine-tuning and test the effect of different \emph{temperatures} on the model's performance.

In the context of large language models, the temperature parameter, $\temp$, sometimes also called the creativity parameter \cite{roemmele2018automated,peeperkorn2024temperature}, controls the randomness of the model's output by scaling the probabilities of the next predicted token when generating traces (or text). 

More formally, we follow the definition given in \cite{peeperkorn2024temperature,ackley1985learning}.
For a given model output logits vector $z\in \mathbb{R} ^n$, we apply the softmax function (Equation \ref{eq:softmax}) with the temperature parameter $\temp$.
% so not a single hyper parameter 
\begin{equation}\label{eq:softmax}
    \textsc{softmax}(z)_i=\frac{
    \exp{(\frac{z_i}{t}}
    )}
    {\sum_j^n \exp{(\frac{z_j}{t})}}
\end{equation}
Where $n$ is the vocabulary size of our model (Equivalent to the number of nodes in the network for which the trace is generated), and $\textsc{softmax}(z)_i$ is the probability for token $i$. The probability distribution for all tokens will increase skewness for lower temperatures $t<1$ or be less skewed for higher temperatures $t>1$. Setting $t=1$ gives the model's default behavior; we will denote this as the default temperature. Typically, the temperature is a real number in the range $t\in [0,2]$. 

It is difficult to predict exactly the effects of different temperatures on the model's output, and the optimal temperature depends on the desired output characteristics. In general, we expect that a lower temperature leads to a model that generates fewer novel sequences, while a higher temperature model generates more. We, therefore, see this as part of the model's fine-tuning process. 
In this paper, we will explore different temperatures when generating novel traces and see the effect of different temperatures on the generated traces for different models.

\subsection{Tokenization}\label{subsec:tokenz}
Tokenization is essential to LLMs. LLMs use tokenization to break down natural human language text into 'tokens', which can then be fed into the model. The tokenization scheme the model utilizes can be on the character, the sub-word or word level, or a different combination of these word structures. In each scheme, A token is assigned a number, and the entire text is transformed into a list of tokens. Tokenization is used to increase the performance of LLMs by reducing the computational load on the model and improving accuracy by indirectly increasing the context length. %However, tokenization can come with a few drawbacks for some tasks. 
The number of possible tokens found in the input sequence is a hyperparameter of the model, and increasing the number of possible tokens will increase the model's complexity by increasing the vocabulary size; thus, this number must be chosen carefully.  
As we have mentioned, in our framework, network traces are formed from a list of the source and destination nodes of the network. Typically, these would be IP addresses. For this preliminary work, we used a straightforward tokenization scheme for our traces. Each ID (source or destination IP, etc.) in the original trace is assigned a number. For the Facebook rack traces, each rack ID in the original trace was reassigned a number to ensure all tokens are consecutive (i.e., from $0$ to the $n-1$ where $n$ is the number of source nodes in Table \ref{tab:traces_data}) The HPC traces were used almost unchanged (from the source in \cite{trace-collection}), with the original ID values reduced by one, that is, from the range $[1,1024]$ they were moved to the range $[0,1023]$. 
We considered a naive approach where all requests (i.e., source and destination ID) are bundled into a single token. However, this would square the vocabulary size (from $n$ to $n^2$) and only double the 'context length.' Furthermore, not all pairs would even be used, and thus, this is very wasteful.    
It is possible to consider different tokenization schemes for network traces, such as bundling longer sequences of IDs in the trace into a single token if they are frequent (much like a several-character word is bundled into a single token), perhaps using byte-pair encoding for tokenization. However, while natural language text has many properties common across all sources, this is likely not true for traffic traces, particularly those of different source networks. Thus, a useful token for one trace might be completely irrelevant for another. Finally, we note that a deeper discussion of tokenization is outside the scope of this paper and will require further research into datacenter traffic traces.

\subsection{Training Setup}\label{subsec:trainsetup}
This preliminary work only evaluates a single model with a single set of hyperparameters used for training. We use a similar set of hyperparameters as the smallest GPT-2 model \cite{GPT22019language}. We used an embedding dimension of $\dd=768$. The number of linear layers $L$ and the number of self-attention heads $H$ were set to $L, H=12$.  Furthermore, the vocabulary size was set as the most significant number of IDs found in the traces, $1024$. Our context length was set to $512$, and the batch size was set to $2^16$. Regarding the meta-embedding parameters, we set the number of segments to $\seg= 24$, and the largest traces ID is $\traceN=7$. The ID of each trace was chosen arbitrarily. Therefore, in our settings, the meta embedding vector has $2\cdot 24 \cdot7=336$ possible tokens.
Traces are divided into a train and validation set, where the validation set is $10\%$ of the tokens. During training, each sequence is randomly sampled from each trace, and the source trace for each sequence is selected randomly and proportionately to the length of the trace; that is, a sequence from a longer trace will be selected more frequently. The training was conducted until the cross entropy validation loss gain was minor.

\subsection{Trace Generation}\label{sec:Tracegeneration}
Let us briefly discuss the parameters we used for trace generation. 
As our result trace set, for each trace from our data set (in Table \ref{tab:traces_data}), we (at least) partially recreate all traces using the following set of parameters. 

We generated Each HPC trace to its original length, or at most $8M$ requests.  While for Facebook, we generated traces that are $2M$ requests long. The actual generated trace lengths are listed in Table \ref{tab:traces_data}.
Each generated trace has five versions, one for each temperature value from the set $\{0.9, 0.95, 1, 1.1, 1.2\} $.
To produce the meta-data embedding, we use the corresponding token ID of each trace, and for the segment tokens, we assume that the length of each segment is the original segment length of the trace. 
We note that the Facebook traces are from a network with a smaller scale; that is, it has a smaller set of IDs. We, therefore, mask the output probability distribution to ensure that no values above the largest ID are generated for those traces. 
Throughout this paper, we note that, unless otherwise specified, every generated trace was created with the default temperature (i.e., $t=1$). When presenting results and comparing traces, all original traces were truncated to be of the same length as their generated counterpart presented in the result.

\section{Evaluation}\label{sec:eval}
\begin{figure*}[h]
  \begin{centering}
   \begin{tabular}{ccccccc}

  \subcaptionbox{CNS}{\includegraphics[width=\matSize\textwidth]{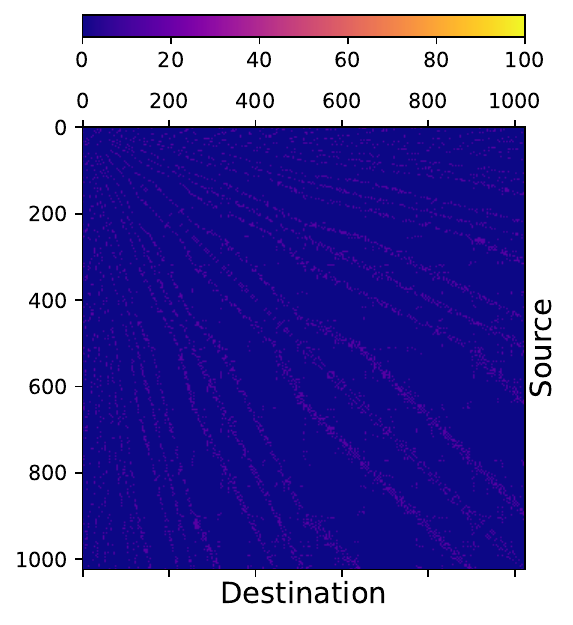}}&
  \subcaptionbox{MultiGrid}{\includegraphics[width=\matSize\textwidth]{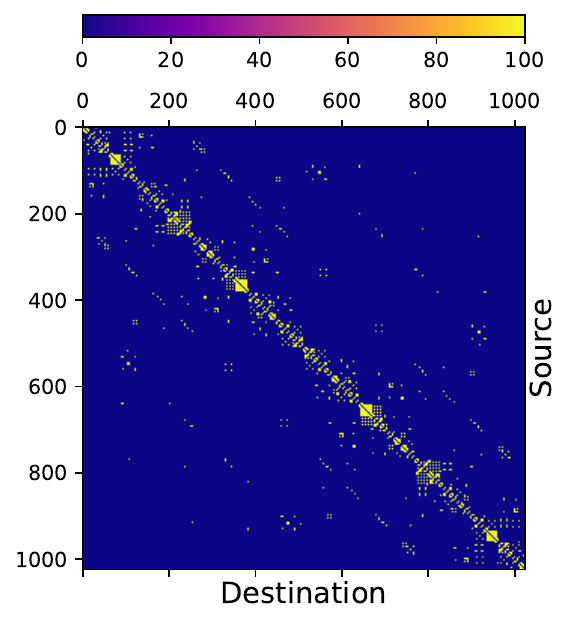}}&
  \subcaptionbox{Mocfe}{\includegraphics[width=\matSize\textwidth]{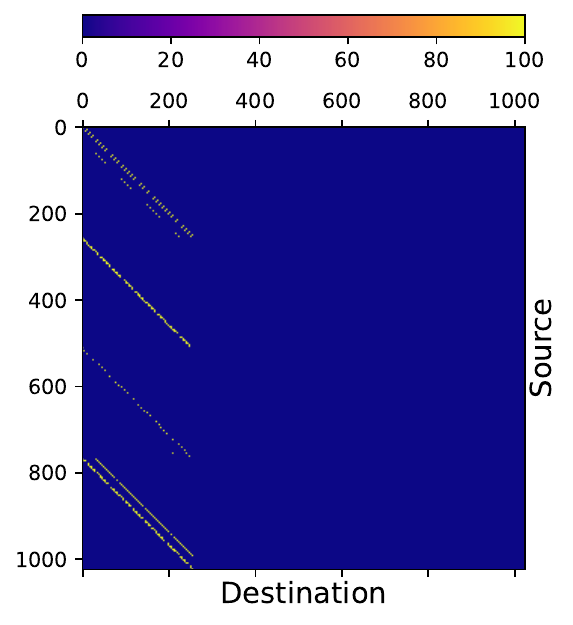}}&
   \subcaptionbox{Nekbone}{\includegraphics[width=\matSize\textwidth]{Figs/Neckbone_orig_matrix_max_100.pdf}} &
  \subcaptionbox{WEB}{\includegraphics[width=\matSize\textwidth]{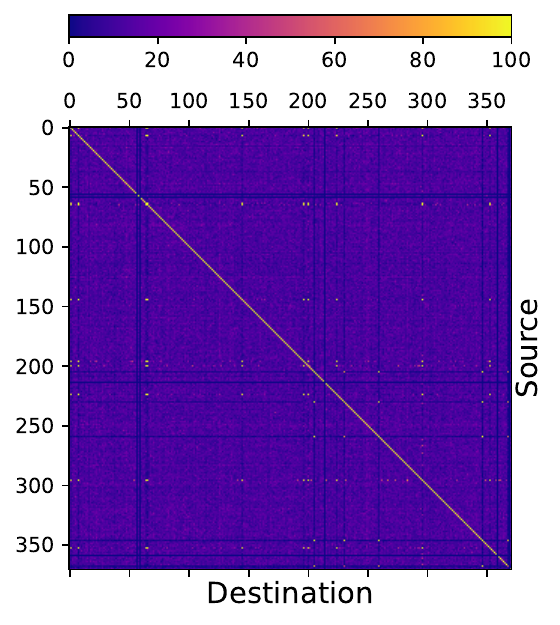}} &
    \subcaptionbox{Hadoop}{\includegraphics[width=\matSize\textwidth]{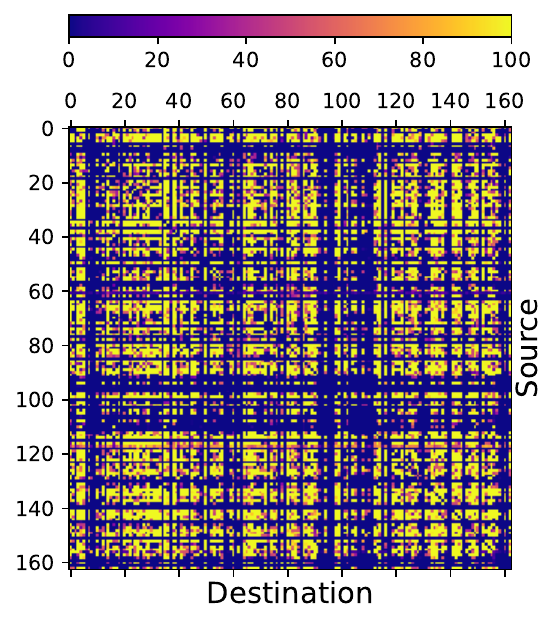}}&
     \subcaptionbox{DB}{\includegraphics[width=\matSize\textwidth]{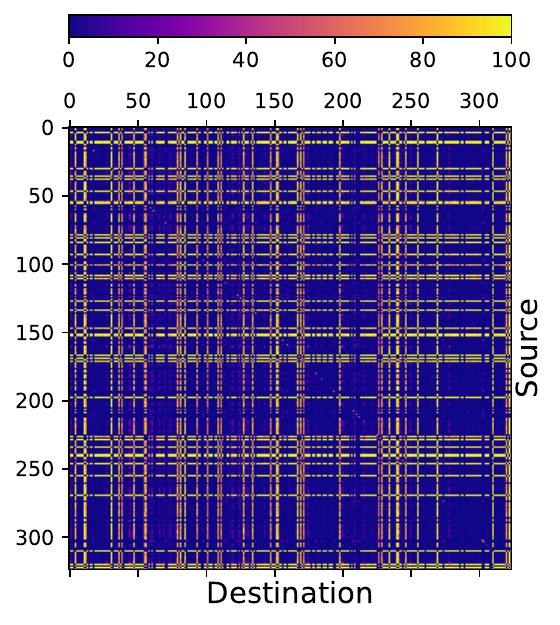}}
     \\
    \subcaptionbox{CNS}{\includegraphics[width=\matSize\textwidth]{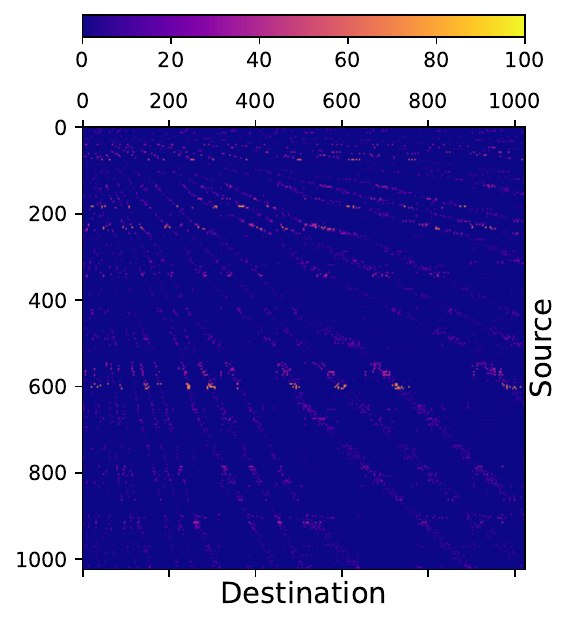}}&
  \subcaptionbox{MultiGrid}{\includegraphics[width=\matSize\textwidth]{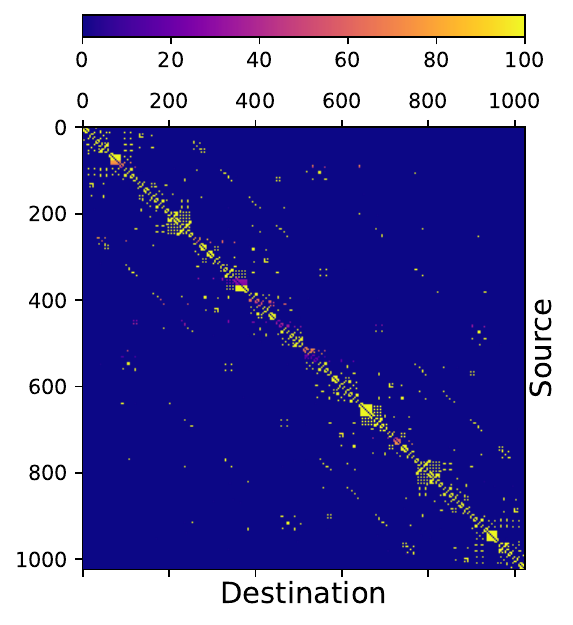}}&
  \subcaptionbox{Mocfe}{\includegraphics[width=\matSize\textwidth]{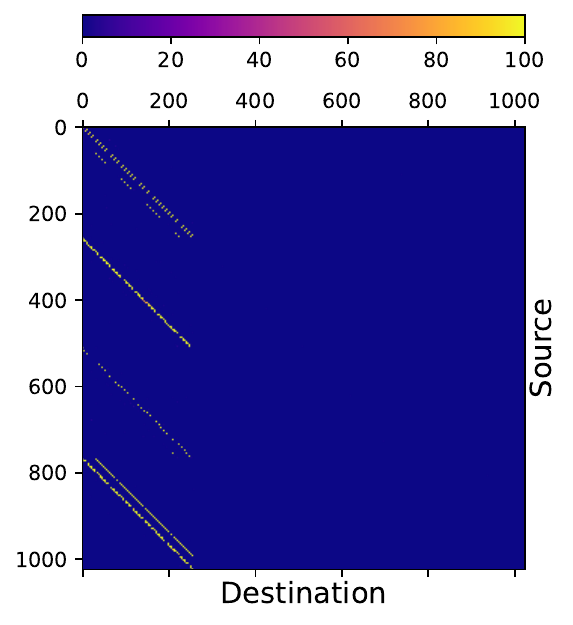}}&
   \subcaptionbox{Neckbone}{\includegraphics[width=\matSize\textwidth]{Figs/Neckbone_model_matrix_max_100.pdf}} &
  \subcaptionbox{WEB}{\includegraphics[width=\matSize\textwidth]{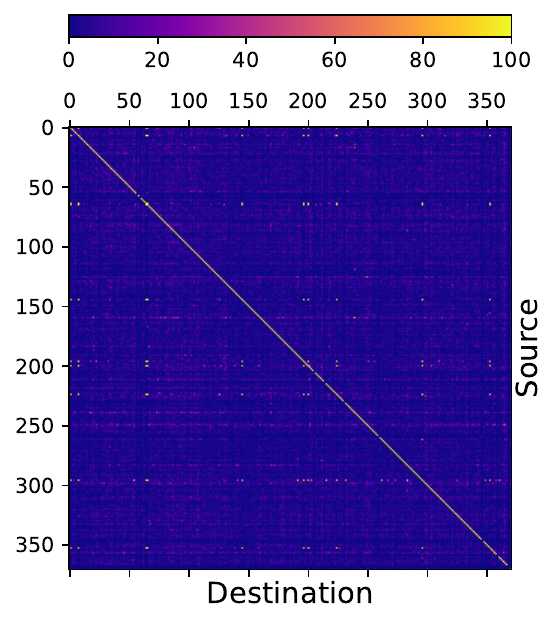}} &
    \subcaptionbox{Hadoop}{\includegraphics[width=\matSize\textwidth]{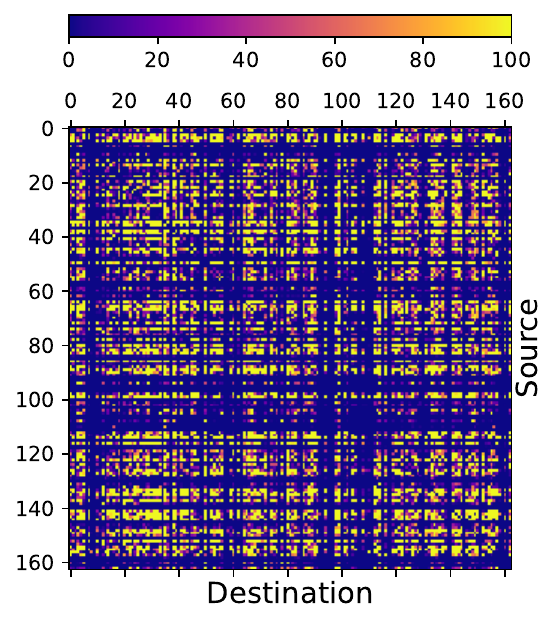}}&
     \subcaptionbox{DB}{\includegraphics[width=\matSize\textwidth]{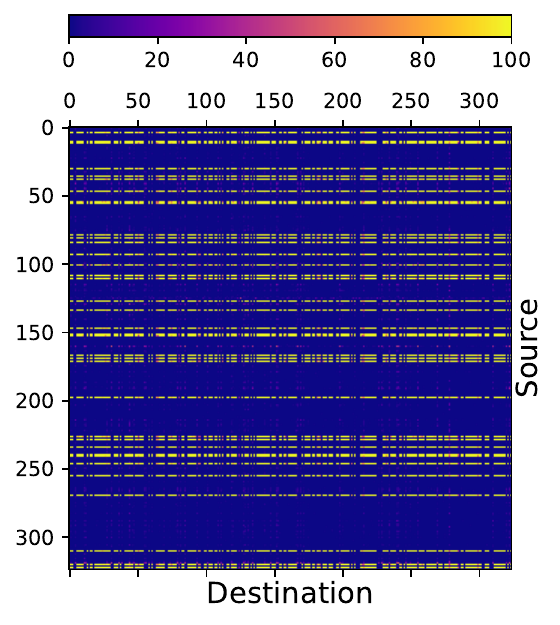}}
\end{tabular}
    \caption{Traffic matrices for several of the communication traces. Colors are scaled individually, and the scale is provided at the top of each matrix. Axes represent source IDs (vertical) and destination IDs (horizontal). 
    }
    \label{fig:allmatrix}
  \end{centering}
\end{figure*}

In this section, we evaluate the fidelity of the traces generated from our \model model to the original set of traces described in Section \ref{subsec:data}.
On every measurement we take on a trace throughout this section, our goal will be to find if traces generated by \model recreate the patterns found in the original traces. We are not seeking an exact match for every measure. Our goal is to generate traces that are similar but different. 
 
We will judge the fidelity of generated traces using four measures. We first look into the traffic matrices (as shortly presented in Section \ref{sec:intro}). We take an information-theoretic approach and analyze the trace complexity of our traces, as discussed in Section \ref{subsec:perlim:comp}. We then explore the burstiness of the generated trace. Finally, we look into the novelty of the generated traces by looking into the ratio of similar n-grams of each trace.
In the appendix, in Section \ref{app:smallScale}, we present a partial result; we test our model's ability to generate a trace representing the traffic for a network with a smaller set of nodes than found in the trace, which it is trying to mimic.

\subsection{Traffic Matrices} 
Traffic traces can have temporal or non-temporal (spatial) aspects \cite{complexity2020}. In this section, we examine the spatial aspect by looking at traffic matrices generated from our generated model. 

Figure \ref{fig:allmatrix} (a)- (g) presents the traffic matrices for our dataset, that is, the original traces, while \ref{fig:allmatrix} (h)- (n) represents the generated counterparts. Each matrix represents an accumulation of all requests in a trace and reveals their spatial structure, as more frequent requests appear as brighter colors. Furthermore, all cells in the matrix were clipped to be at a size of at most $100$; this allows for better contrast between cells.    
In Figure \ref{fig:allmatrix} (a)- (g), we can observe that each of the traffic matrices, and in particular, the HPC matrices, present a clear and unique pattern.
Looking at Figure \ref{fig:allmatrix} (h)-(n), we observe that, indeed, our model recreates the general pattern in the traffic matrix. Looking closely, the major difference in any matrix is the frequency of some areas or cells; this is apparent in \ref{fig:allmatrix} (a) $\&$ (h) for the CNS trace and the Facebook matrices. 
More generally, the model seems to more easily recreate traces from HPC. In particular, the DB traces seem to miss some perpendicular lines. We conjecture that the Facebook set is more random and holds less structure and, therefore, might require a larger model. In the scope of this paper, we examine the potential of setting a different temperature for the model to improve the result.

\begin{figure}[h]
  \begin{centering}
   \begin{tabular}{ccc}

  \subcaptionbox{Original}{\includegraphics[width=0.14\textwidth]{Figs/Web_orig_matrix_max_100.pdf}} &
  
    \subcaptionbox{t:0.9}{\includegraphics[width=0.14\textwidth]{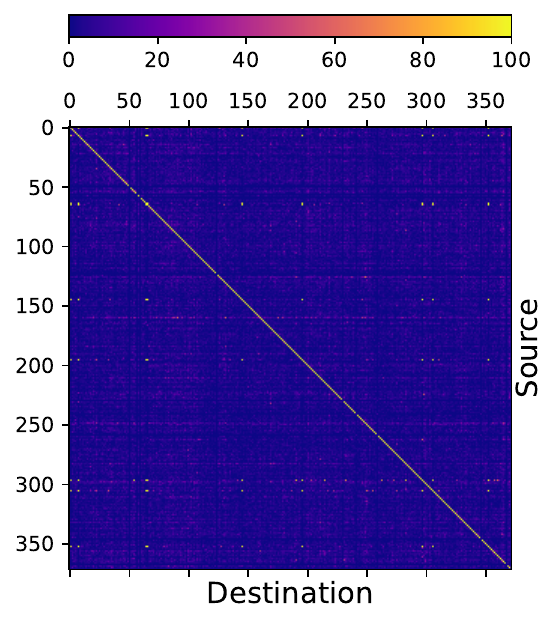}}&
     \subcaptionbox{t:1.2}{\includegraphics[width=0.14\textwidth]{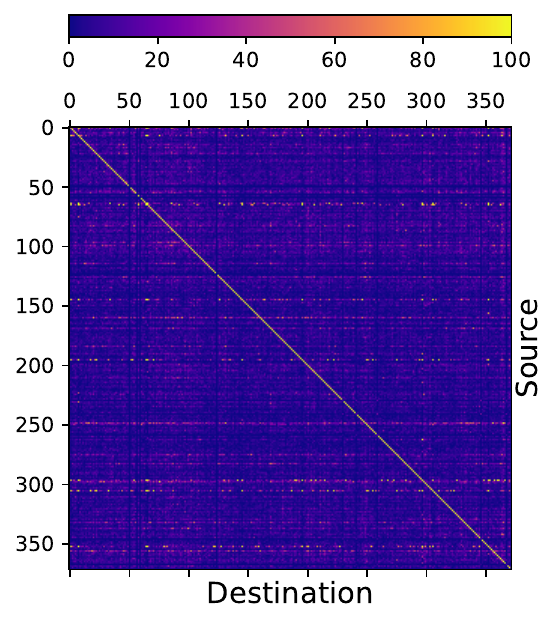}}

\end{tabular}
    \caption{Original WEB trace traffic matrix, along with generated traces using two different temperatures.
    }
    \label{fig:tempmatrix}
  \end{centering}
\end{figure}

In Figure \ref{fig:tempmatrix}, we see three different traffic matrices for the DB trace and two generated counterparts of two different temperatures. In Figure \ref{fig:tempmatrix} (b), the generation temperature was set to $0.9$; in Figure \ref{fig:tempmatrix} (c), the generation temperature was set to $1.2$. Figure \ref{fig:tempmatrix} (a) shows the original traffic matrix for reference. 
The pattern of the original trace is not very specific. It can be broadly described with an active diagonal and a few active hot spots, and the rest of the traffic is more uniformly distributed except for a few nodes. The low-temperature matrix in \ref{fig:tempmatrix} (b) shows a good similarity regarding the most active features, such as the diagonal and the hot spots. However, the rest of the traffic seems to be less uniform. For \ref{fig:tempmatrix} (c), it also shows a good reconstruction of the most active features, but the rest of the matrix seems more uniform and closer to the original. We can tentatively conclude that, in this case, a trace generated using a higher temperature setting better recreated the original.

We note that recreating a traffic matrix is simple on its own. We merely need to sample each trace from the requested distribution. 
We offer this as a basic validation of our models' performance. It allows us to be sufficiently sure that the spatial structure of the distribution created by the model is of a similar nature to the original trace it mimics. 
In conclusion, we see that our model can recreate the general pattern found in the original traffic trace. This is also consistent with a model generating a trace using different temperatures from the tested model; fine-tuning the model with different temperatures may help create a more accurate trace. 
However, recreating a trace requires far more than generating a similar traffic matrix. In the following chapters, we will focus more on the temporal aspect of the trace and see if our model can generate a similar trace in the temporal sense; we will show how the nature of the GPT architecture can generate a trace of a similar temporal pattern.

\begin{figure*}[t]
  \begin{centering}
   \begin{tabular}{cc}
  \subcaptionbox{HPC}{\includegraphics[width=0.4\textwidth]{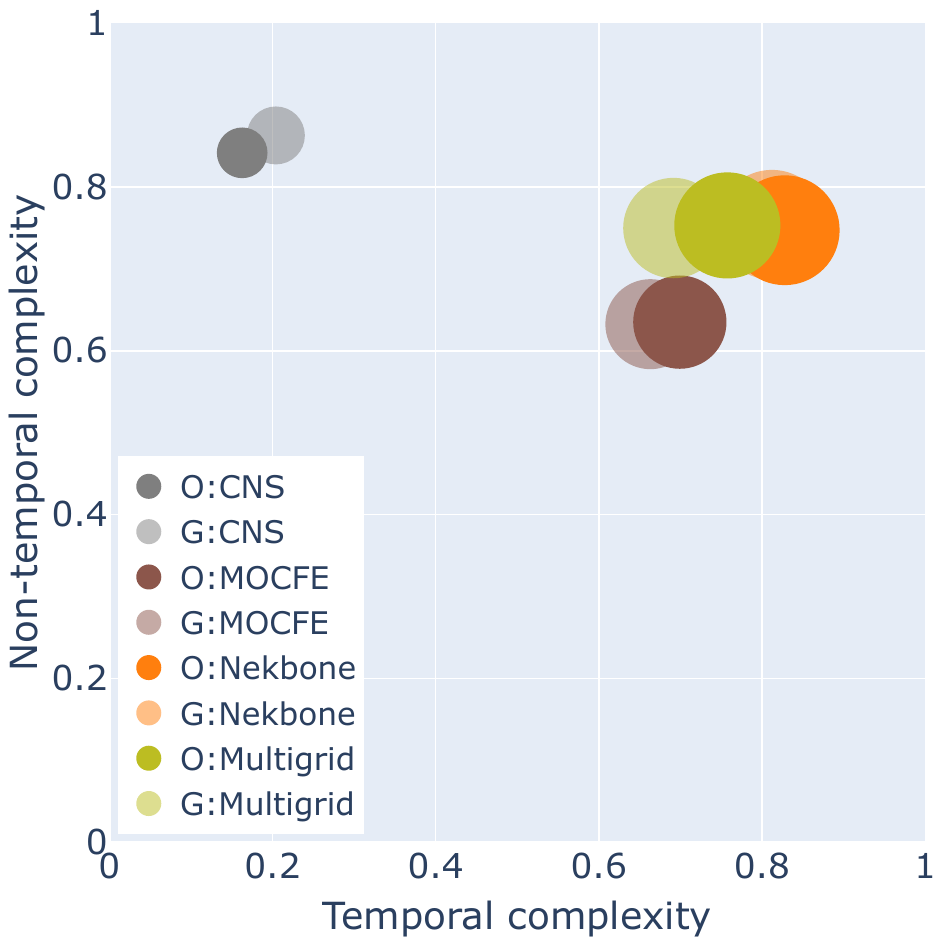}} &
    \subcaptionbox{Facebook}{\includegraphics[width=0.4
    \textwidth]{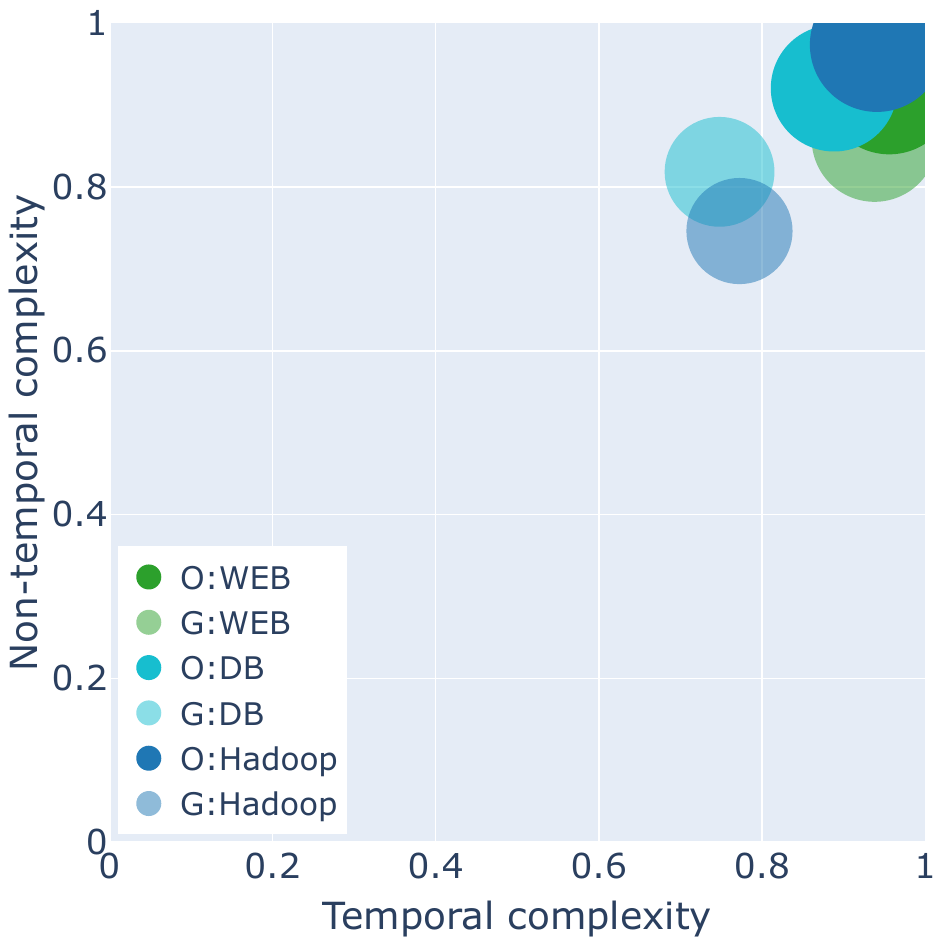}}
\end{tabular}
    \caption{The complexity map of seven real traces and their generated counterparts. An opaque circle designates original ($O$) traces, while generated ($G$)  traces have a more transparent circle. (a) Represents traces from the HPC set. (b) Represents traces from the Facebook set. 
    }
    \label{fig:compmapMain}
  \end{centering}
\end{figure*}
\subsection{Trace Complexity}

In this section, we explore trace complexity, as described in Section \ref{subsec:perlim:comp}. We implemented the process found in the original paper \cite{complexity2020} and calculated complexity values for all traces.
We believe that if the traces generated from our \model model are indeed faithful to the original traces, they will have a similar \emph{complexity profile}, i.e., similar values of both temporal and non-temporal complexity. In this section, we will discuss this hypothesis. 
In Figure \ref{fig:compmapMain} (a) and (b), we see the complexity maps for the HPC and Facebook trace sets, respectively. Each trace is paired with a generated trace from our model, marked by a transparent circle of the same color. We can observe that our original trace set is diverse in terms of its complexity values, particularly on the temporal scale, with traces having temporal complexity values ranging from roughly $0.2$ to near $1$.  
Regarding the generated traces, in \ref{fig:compmapMain} (a), we observe that each circle for each generated trace overlaps with its original trace circle; this means that both traces have a similar complexity profile. We note that the generated Neckbone trace is nearly identical. Looking at each complexity dimension (the $x$ or $y$ axis) separately, we see that the generated trace follows the non-temporal structure of their original counterparts closely, as the non-temporal complexity is nearly identical on all traces, with the CNS trace being somewhat of an outlier; this is concordant with our previous result looking at the traffic matrices. We see that the generated traces also contain a fairly similar quantity of temporal structure, which is the main difference between the traces.   
In \ref{fig:compmapMain} (b), we see that only the generated WEB trace overlaps with its original counterpart, while the others are further apart, with the Hadoop trace being the most separated from its generated counterpart. We conjecture that these traces are more difficult to mimic as they are more random, as evident by their larger overall complexity. 

Let us explore if setting a different temperature during trace generation could help us achieve a more accurate complexity profile.  

\begin{figure}[h]
  \begin{centering}
  
  \includegraphics[width=0.4\textwidth]{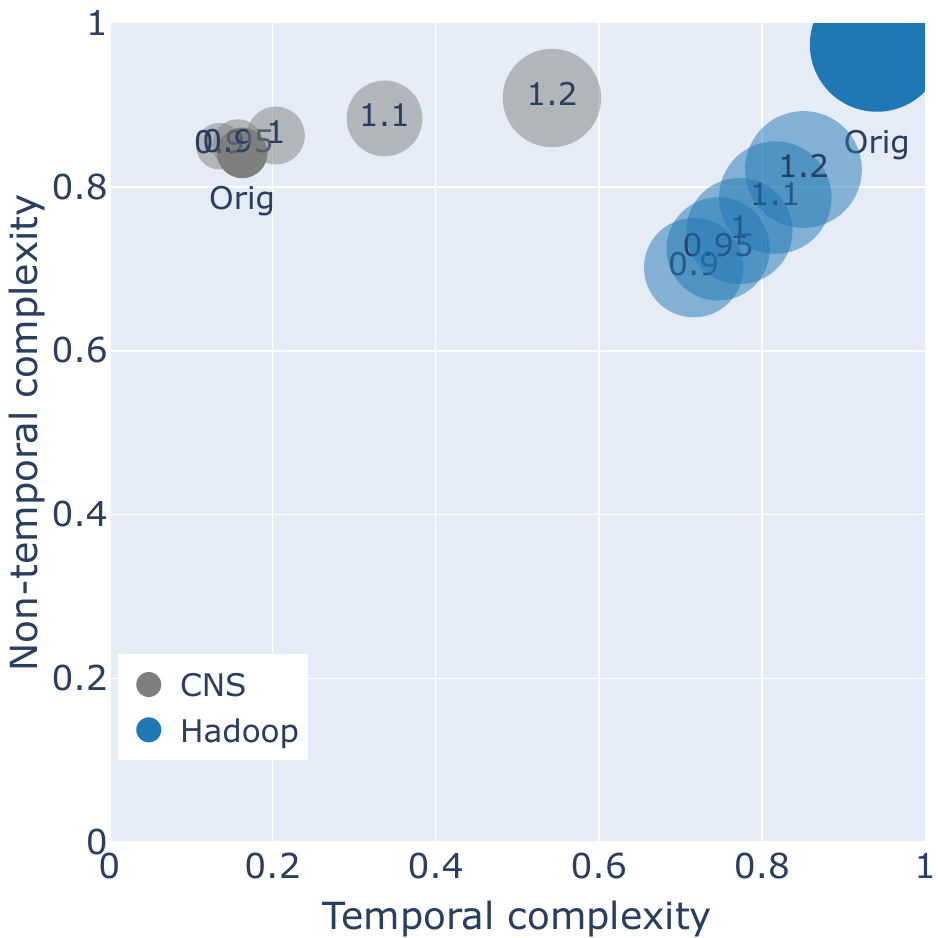}

    \caption{The complexity map for the HPC CNS and Facebook Hadoop traces and their generated counterparts for five different temperature settings. An opaque circle designates original traces, while generated traces have a more transparent circle.   
    }
    \label{fig:compMapTemp}
  \end{centering}
\end{figure}
\begin{figure}[h]
  \begin{centering}
  
  \includegraphics[width=0.46\textwidth]{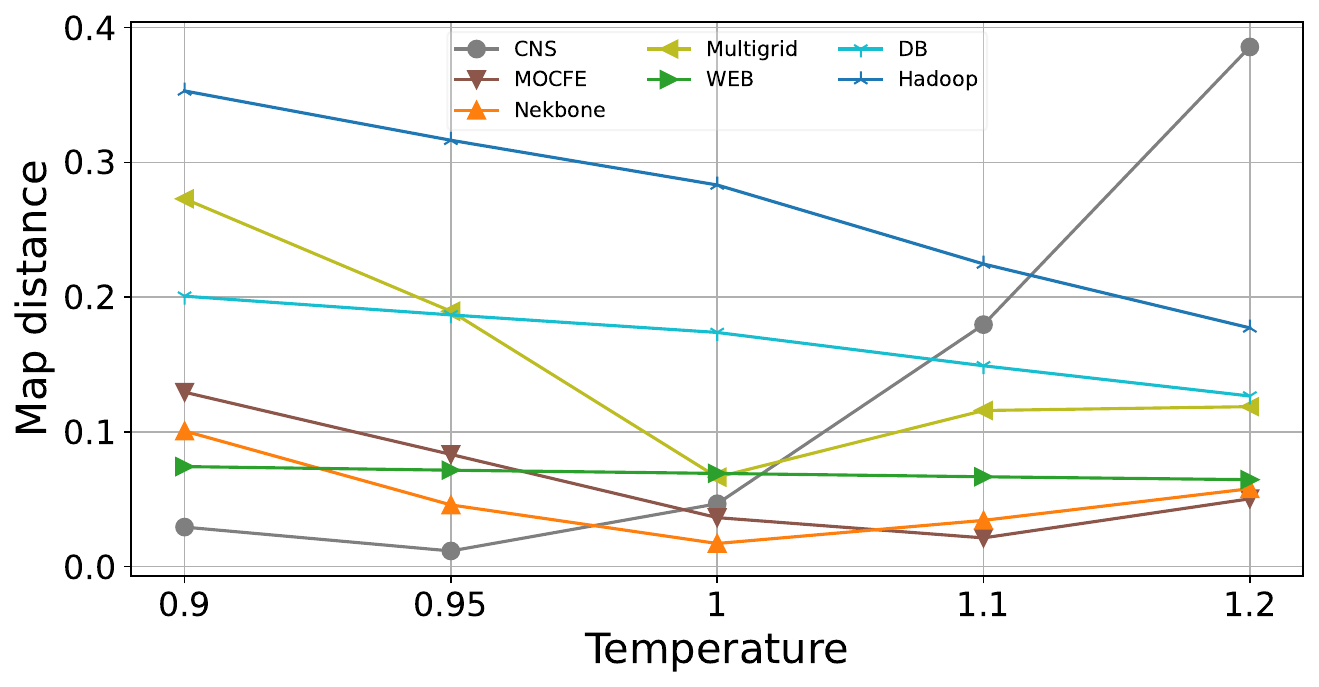}

    \caption{The Euclidean distance (in the complexity map) of the original traces to generated traces with five different temperatures.   
    }
    \label{fig:comlexityDist}
  \end{centering}
\end{figure}

In Figure \ref{fig:compMapTemp}, we see the complexity map for the two most 'misaligned' traces from each trace set in Figure \ref{fig:compmapMain}; these are the HPC CNS trace and the Facebook Hadoop trace. The map in Figure \ref{fig:compMapTemp} also contains the results of five generated traces with different temperatures from the set $[0.9,0.95,1,1.1,1.2]$, each generated trace is marked by the temperature used to generate it, while the original traces are denoted by 'Orig.' 
Increasing the temperature increases the randomness of the model, thus we expect that increasing temperature will increase the trace complexity of the generated traces and vise versa.
Indeed, this is evident in the map as each generated trace moves towards maximal complexity at the upper right corner (at point $(1,1)$) with increasing temperatures. Regarding the generated traces, for the CNS trace, we can see that a trace generated at a temperature of $0.95$ better matches the complexity profile of the original CNS trace than a trace generated at a temperature of $1.0$. For the Hadoop trace, we notice that greater temperature improves the complexity profile to make it closer to the original Hadoop trace. 
The complexity map allows us to visualize the complexity profile of different traces on the same graph. However, to judge the similarity of generated traces to the original in a more precise manner, we take the Euclidean distance between the original and generated complexity profiles. That is, we measure the distance between the Cartesian coordinates of both traces. 
Figure \ref{fig:comlexityDist} shows the distance between each of our seven traces and their generated counterparts for a set of five temperatures. We can see that different traces have different temperatures, which minimize the distance between generated and original traces. For example, for the trace CNS, this is at 0.95; for MOCFE, it's at $1.1$, while for the DB and Hadoop traces, we likely may need to increase the range further to find an optimal temperature.  

\subsection{Traffic Bursts}
\begin{figure*}[h]
  \begin{centering}
   \begin{tabular}{ccc}
  \subcaptionbox{Multigrid}{\includegraphics[width=0.31\textwidth]{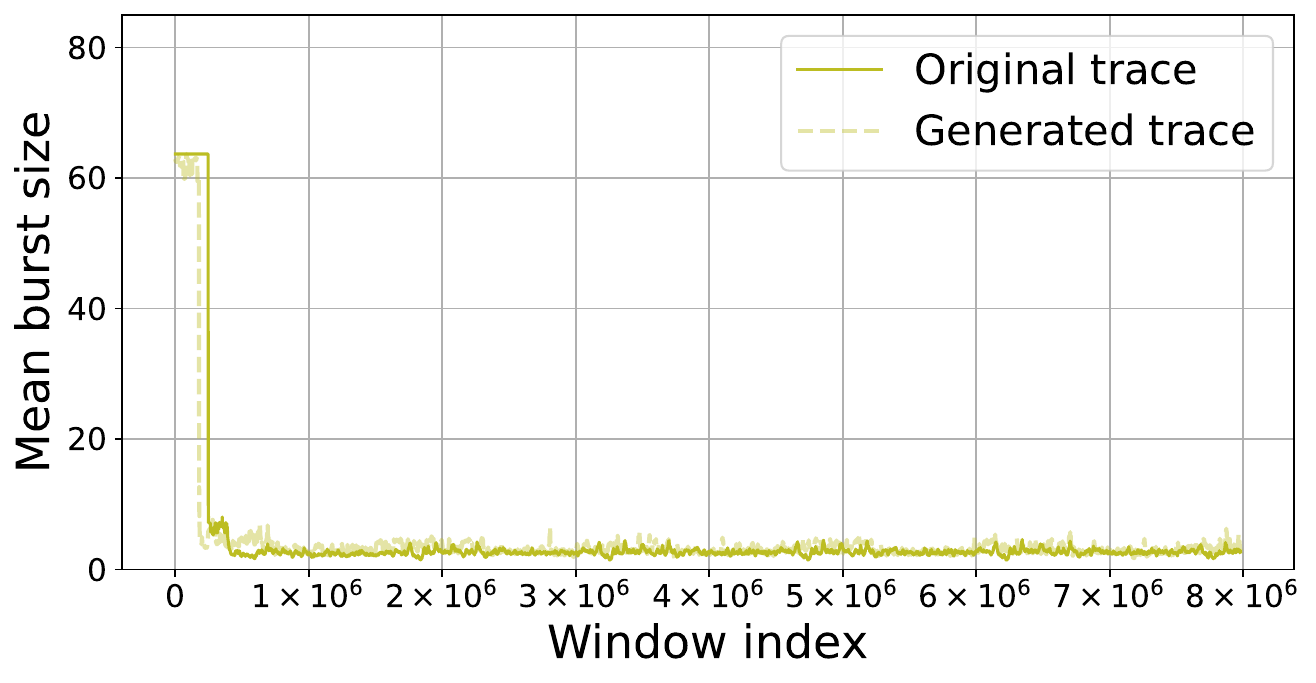}} &
    \subcaptionbox{Nekbone}{\includegraphics[width=0.31\textwidth]{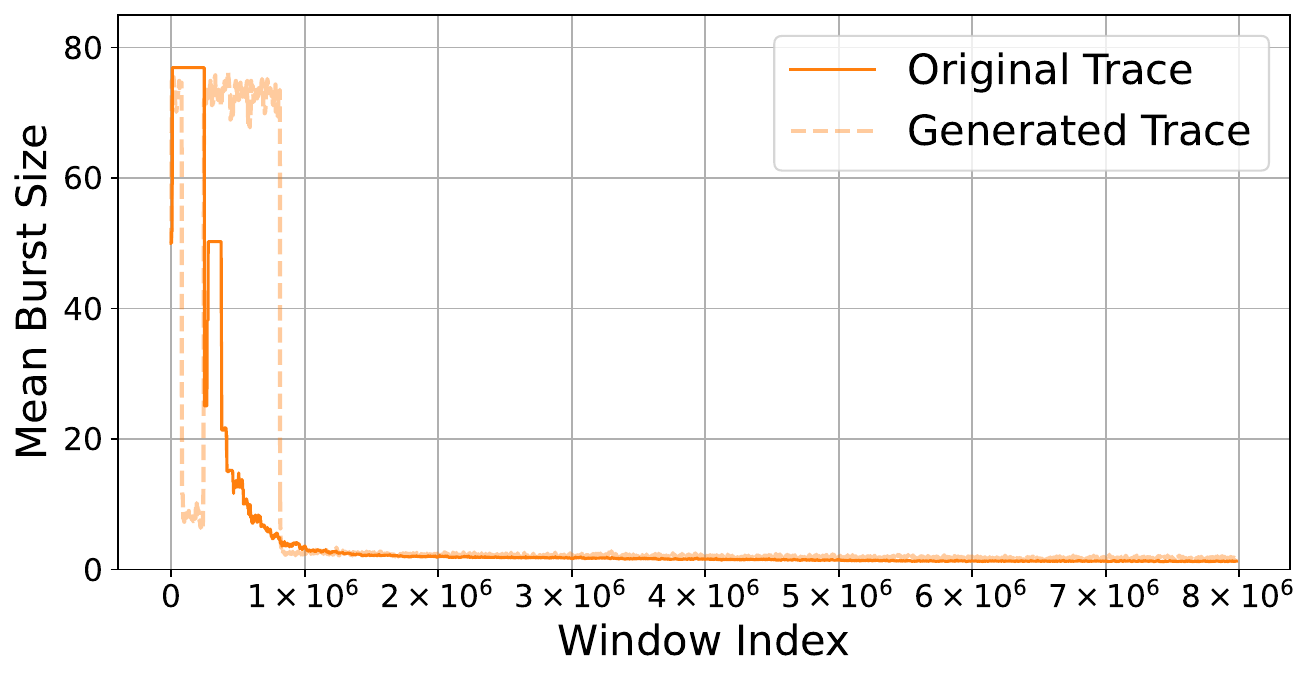}}&
     \subcaptionbox{DB}{\includegraphics[width=0.31\textwidth]{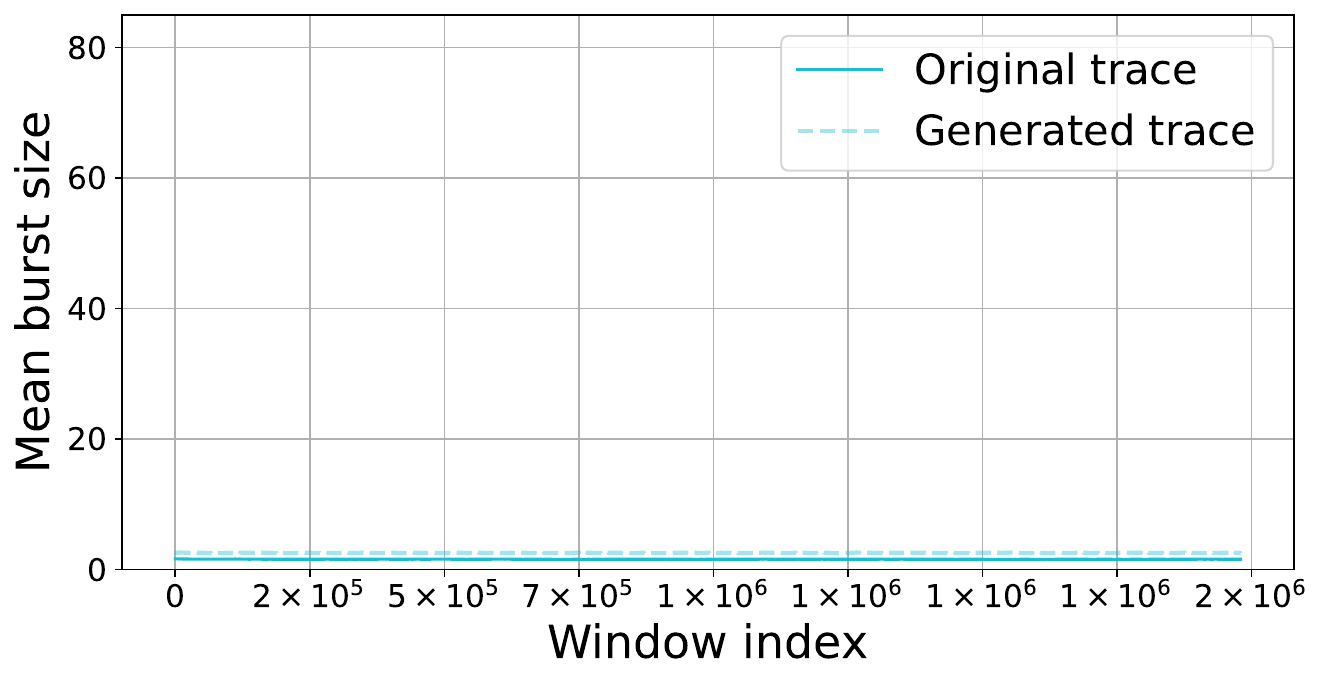}}
\end{tabular}
    \caption{The mean burst size for every overlapping window of size $w=20k$ on three traces. The x-axis reflects the index where each window starts.
    }
    \label{fig:meanBursts}
  \end{centering}
\end{figure*}

\begin{figure*}[h]
  \begin{centering}
   \begin{tabular}{ccc}
 
     \subcaptionbox{Multigrid}{\includegraphics[width=.31\textwidth]{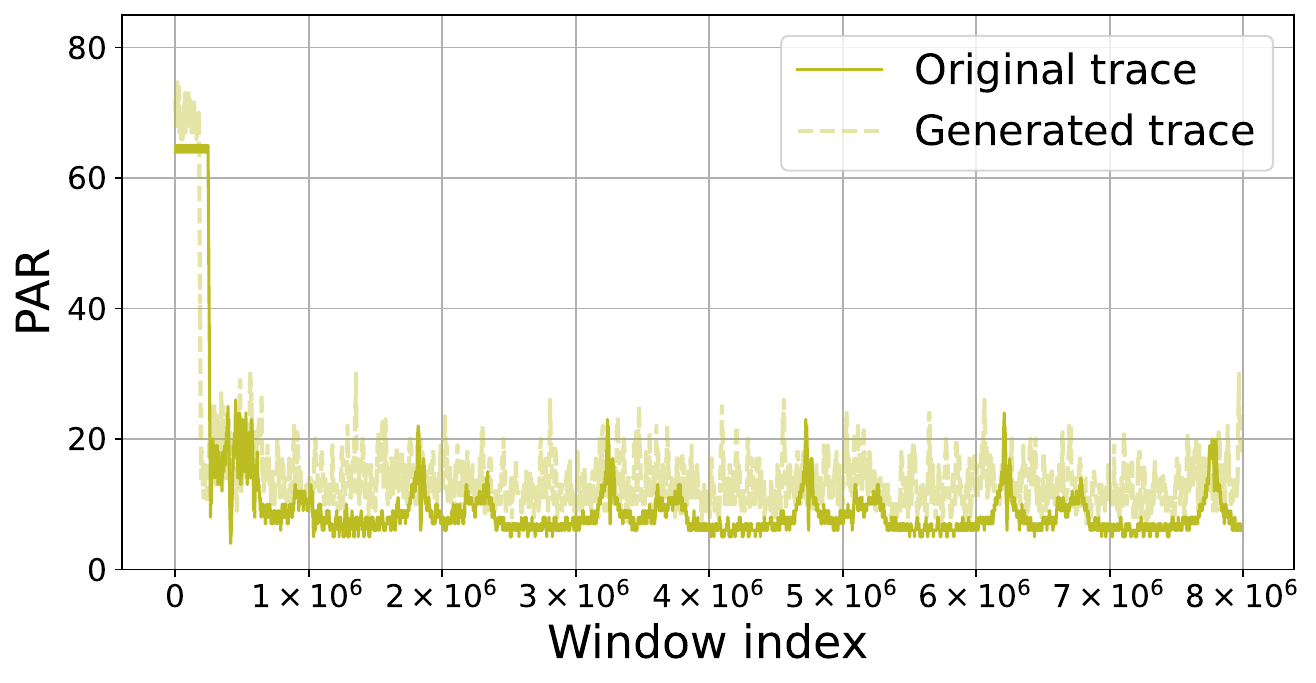}}&
    
    \subcaptionbox{Nekbone}{\includegraphics[width=0.31\textwidth]{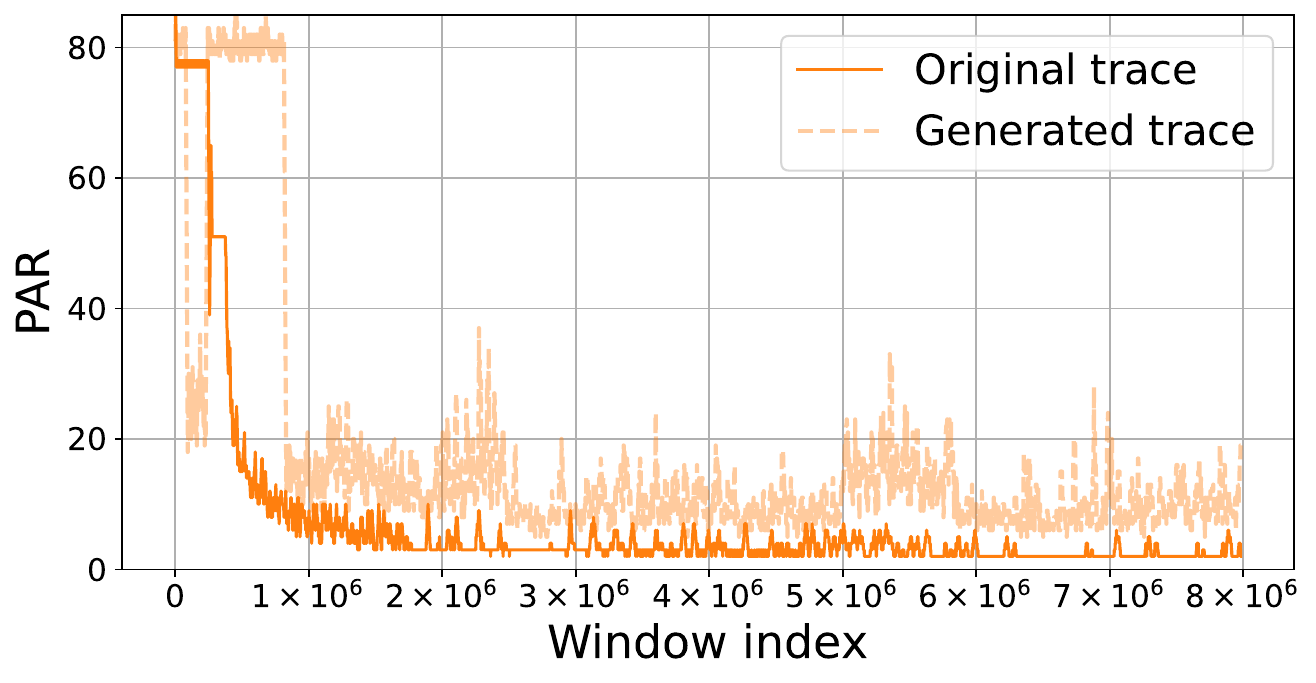}}&
    
  \subcaptionbox{DB}{\includegraphics[width=0.31\textwidth]{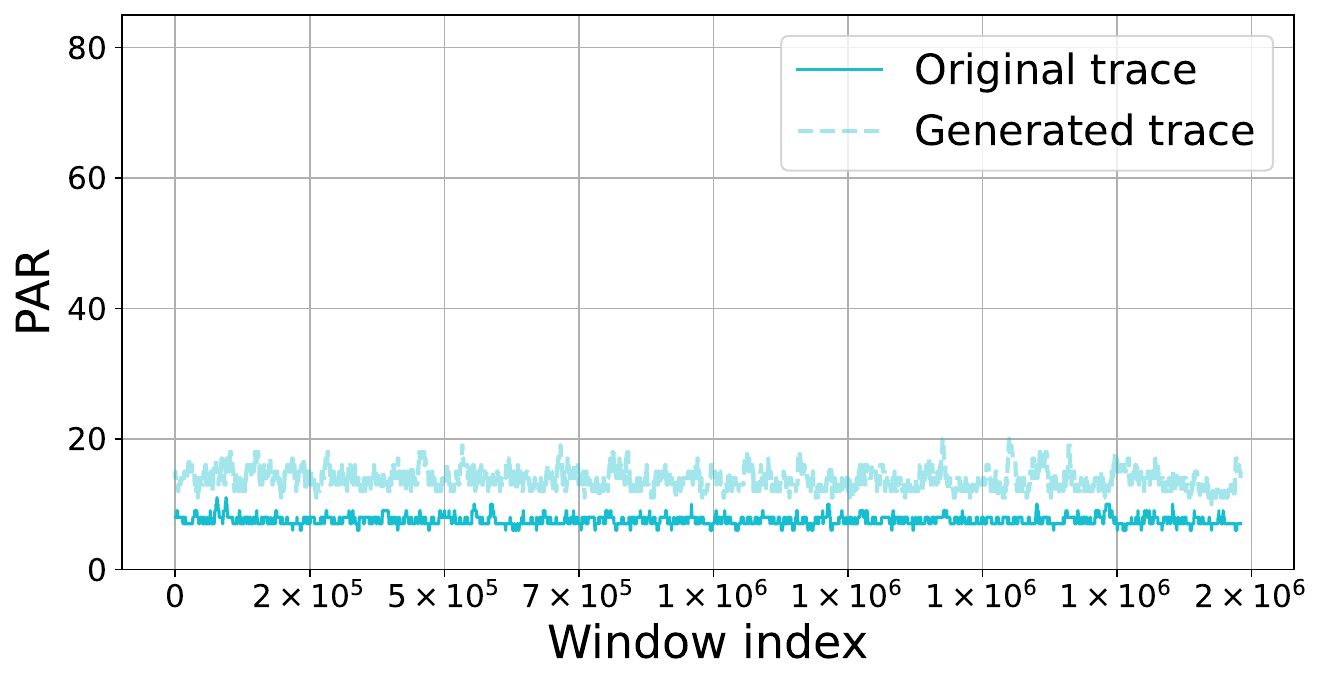}} 
\end{tabular}
    \caption{The burst PAR for every overlapping window of size $w=20k$ on three traces. The x-axis reflects the index where each window starts.
    }
    \label{fig:PAR}
  \end{centering}
\end{figure*}

While trace complexity looks at the fidelity of traces from the point of view of information theory, in this section, we take a more classical network approach and explore traffic bursts.
A very commonly researched topic for network planners is the bursty behavior of the traffic.
Generally, a burst can be defined as a sequence of requests arriving during a short time period, above the typical traffic behavior \cite{zhang2017high}. Traces commonly have an explicit temporal measure (i.e., a timestamp in seconds); however, our traces are more simplistic. Therefore, in our context, we measure our bursts over a window $W$ of $w$ requests; that is, a window $W$ is part of $\sigma$ starting at some index $i$ and ending at $i+w$, $W=\sigma[i,i+w]$. Non-bursty traces will have all possible request types (i.e., source-destination pairs $(s,d)$ ) uniformly distributed through the window, while bursty traces will have only a few request types during the window. For each window, the burst $B_W(s,d)$ of a request $(s,d)$ is the number of times $(s,d)$ appeared in the window sequence $W$.
Let us denote the mean burst size of the window $W$ as $MB(W)$. This can be calculated by counting the unique requests in $W$ and dividing by the window size $w$.  More formally,
\begin{align}
   \textsc{MB}(W)= \sum _{  \forall (s,d)\in W} \frac{1}{w}.
\end{align}
% is is For each trace window, we count the occurrences of each request type and get the mean burst size.
Another interesting measure we would like to test is the ratio between the largest burst in the window $W$ to the mean burst size of the window $\textsc{MB}(W)$, that is, the peak-to-average ratio (PAR). More formally,
\begin{align}
    PAR(W) = \frac{max(B_W(s,d))}{\textsc{MB}(W)}.
\end{align}
Where $max(B_W(s,d))$ is the largest burst in $W$.
The mean burst size serves as an indicator of the uniformity of traffic over time. For each window, a low value means that many pairs communicate during each window, while a high value means only a few communicate. The PAR of bursts provides insight into the magnitude of bursts within the traffic, highlighting the gap between peak and average traffic levels \cite{PARBursts}. A high PAR would show that the traffic is indeed bursty and that, at least a certain pair is very active in a specific window.
In this evaluation, we compare the mean burst size and the PAR of burst and their change over time and examine whether traces generated from the \model show similar bursty behavior to their original counterparts.  
For our \model model to be useful, we believe that generated traces should follow a similar but not identical pattern of bursts as the original trace.

We study three representative examples. Figures \ref{fig:meanBursts} and \ref{fig:PAR}, presents the mean burst size and PAR results for three traces, MultiGrid, Nekbone, and DB, and their generated counterparts on a single window size, $w=20k$. 

Figure \ref{fig:meanBursts} shows the mean burst size for every window in the trace; that is, for every index $x$ in the $X$ axis, we calculated the mean burst size of the window starting at index $x$ in the trace $\sigma$, i.e., $W= \sigma[x,x+w]$. 
Figure \ref{fig:meanBursts} (a) for the MultiGrid trace shows a simple pattern similar to a Heaviside step function. The mean burst size is roughly $60$ for the first few hundred thousand requests, and it then goes to below $10$ requests until the end of the trace with small variations. The generated trace has a very similar pattern but with more variability. Looking more closely, both traces present a small bump at around $0.5M$ to $1.0M$ requests, but the generated trace shows this at a slightly different location.   Figure \ref{fig:meanBursts} (b) for the Nekbone trace shows a similar pattern to MultiGrid overall. However, the initial $1.0M$ requests show a longer bursty period with more variation with two 'plateaus' in this phase for about $80$ and $50$ requests. The generated traces also show two plateaus; however, both are at about $80$ requests and show more variation. Both show a similar for the trace past $1.0M$ requests.  
In \ref{fig:meanBursts} (c), we see that the DB trace shows a very shallow mean. This means that requests are mostly uniformly distributed. However, the generated trace has a slightly larger average than the original DB trace. 

\begin{figure}[h]
  \begin{centering}
  
  \includegraphics[width=0.43\textwidth]{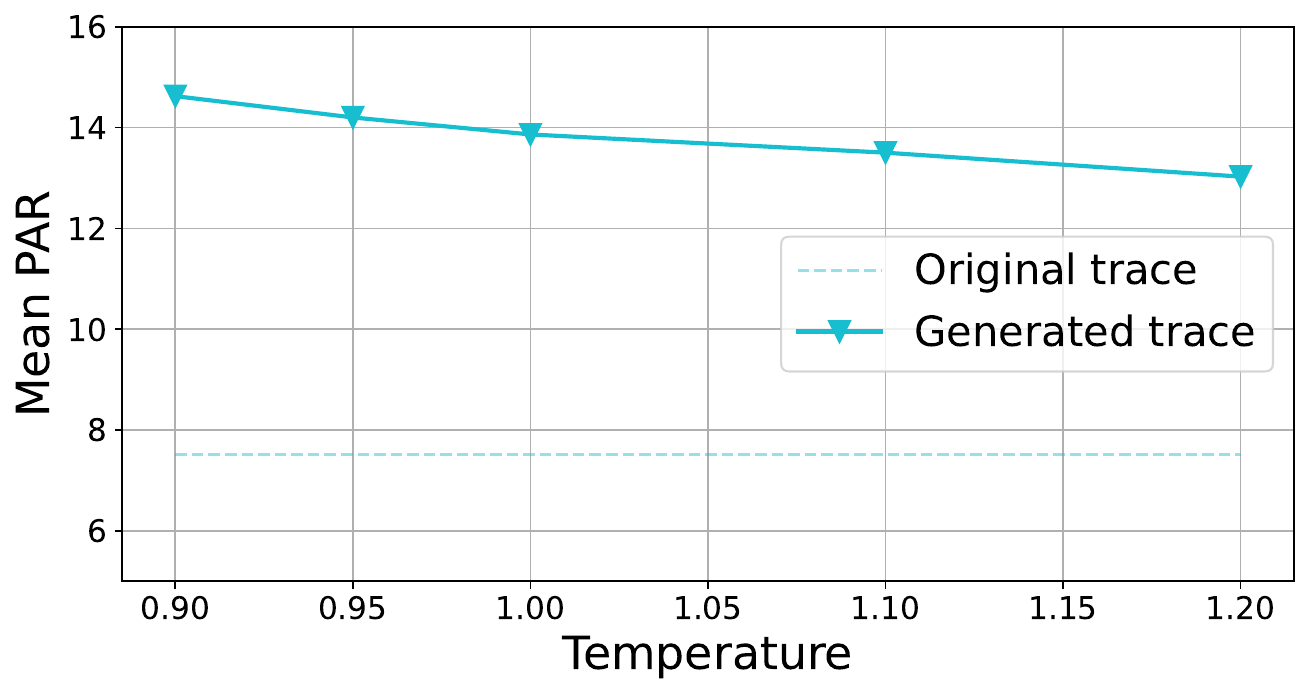}

    \caption{The mean burst PAR value for the generated DB trace against model temperature. The result of the original trace is shown as a reference.   
    }
    \label{fig:burstToTemp}
  \end{centering}
\end{figure}

Figure \ref{fig:PAR} shows the PAR of bursts for every window in the trace as in Figure \ref{fig:meanBursts}. We see that for each trace, the pattern of bursts is generally similar to Figure \ref{fig:meanBursts}, with both the MultiGrid trace and the Nekbone trace showing a pattern of high plateaus in the first $1.0M$ requests and the DB trace being nearly featureless. However, Figure \ref{fig:PAR} (a) shows an interesting periodicity for the MultiGrid trace which is not exactly replicated in the generated counterpart, however, it seems that the range of burst PAR is similar in both case through the trace. Figure \ref{fig:PAR} (b) shows that the generated trace for the Nekbone shows a greater variability and size than the original trace. However, it is possible to see that both traces are more variable in the first part of the trace. Figure \ref{fig:PAR} (c)  for the DB trace shows that the generated trace has a higher burst PAR, but it is otherwise similar to the original trace.
We note that a higher burst PAR should mean lower temporal complexity (as seen in Figure \ref{fig:compmapMain}). We can increase temperature to increase complexity and perhaps reduce burst size. Figure \ref{fig:burstToTemp} shows the mean burst PAR of the generated DB trace plotted against the model's temperature. In the tested temperature range (from $0.9$ to $1.2$), the mean PAR has decreased from $15$ to about $13$. While this is still not a perfect fit to the original trace's value of $8$, we again show that choosing a different temperature for the model may increase performance. 

To conclude, Figures \ref{fig:PAR} and \ref{fig:compmapMain} show that our \model model can recreate traces with a similar level of burst PAR and mean bursts. More importantly, our generated trace shows that the generated trace can replicate at least some nontrivial structures in the burstiness pattern of our tested network traces. We note that changing the model's temperature may benefit some traces. We believe that this evaluation shows the importance of the meta-data vector (as described in Section \ref{sec:Architecture}). Our model recreates some patterns seen in the burst pattern of the HPC traces using information gained from the segment vector. Namely, to recreate the large bursts seen at the beginning of the HPC traces, not at a random time, but at the correct time, the model likely used the segments tokens. However, an in-depth examination of the effects of the segment vector is beyond this parliamentary work's scope.

\subsection{Novelty of Generated Traces}\label{subsec:novelty}

\begin{figure}[h]
  \begin{centering}
   \begin{tabular}{c} 
  \subcaptionbox{HPC}{\includegraphics[width=0.43\textwidth]{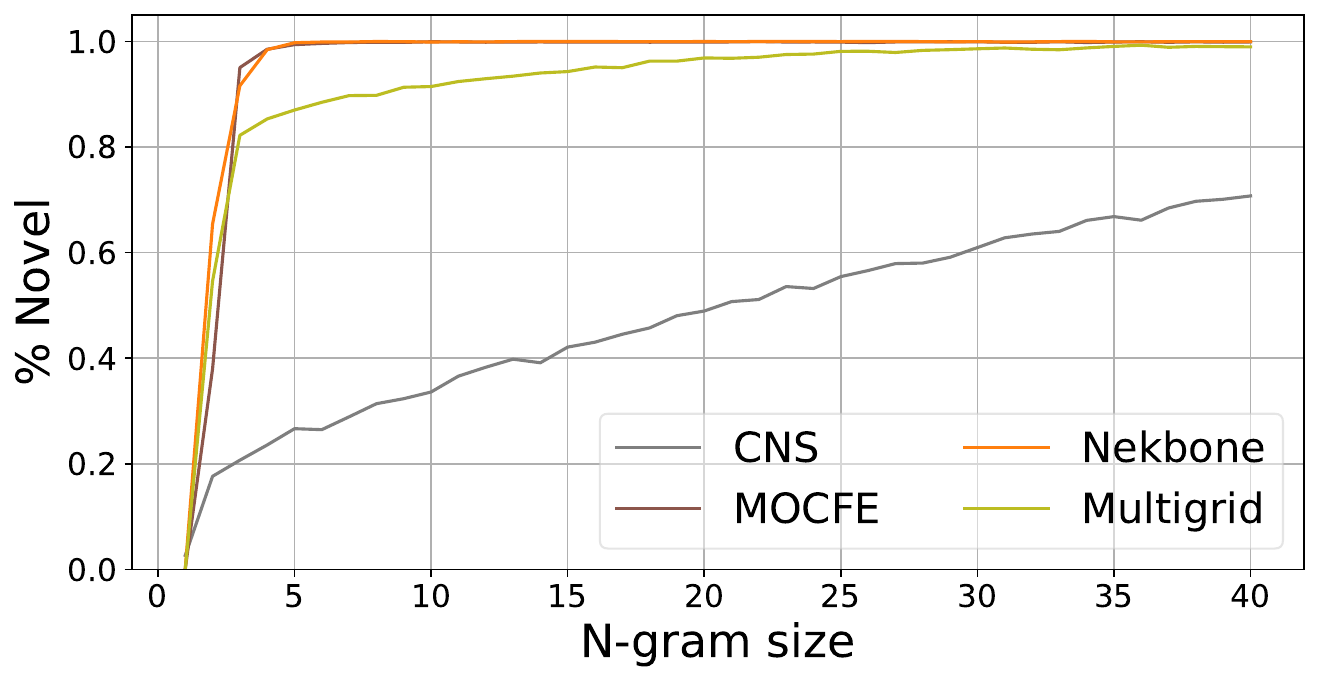}} \\
  
    \subcaptionbox{Facebook}{\includegraphics[width=0.43\textwidth]{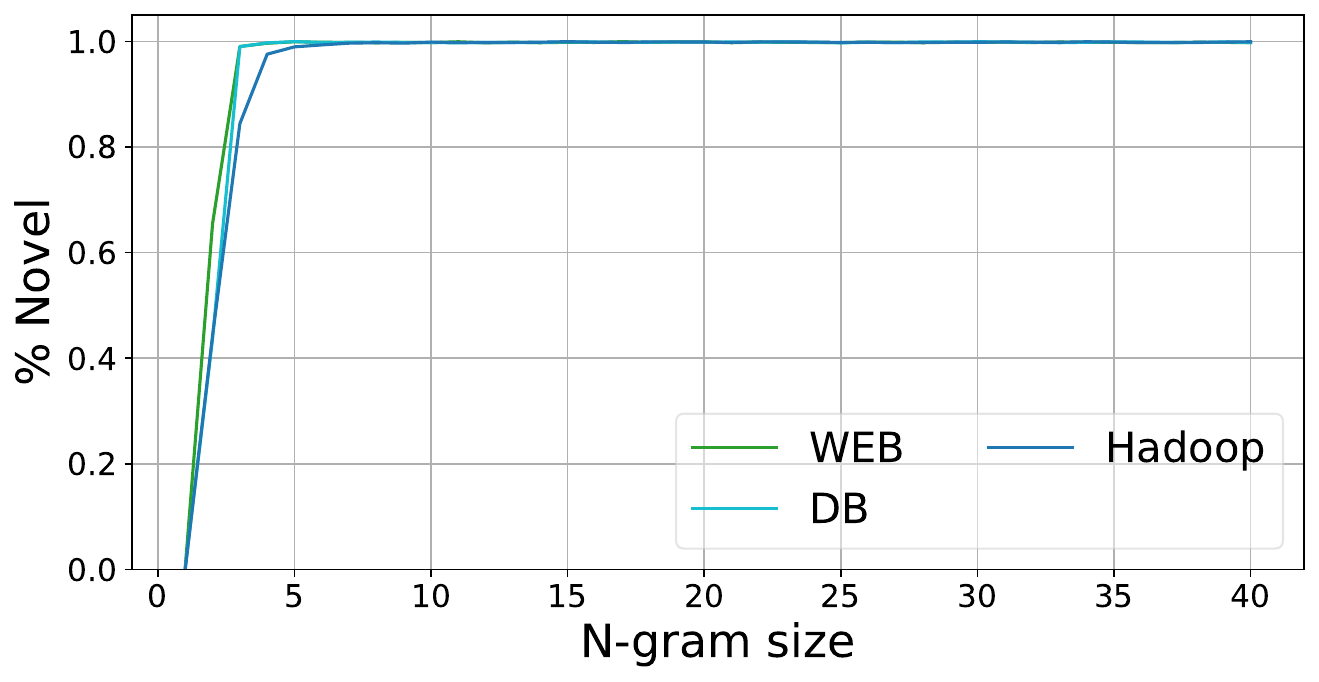}}
\end{tabular}
    \caption{ Proportion of n-grams that are novel against n-gram size between the original and generated trace in two domains (a) for HPC,  (b) Facebook.}
    \label{fig:novelNgram}
  \end{centering}
\end{figure}
As mentioned, our goal when generating new traces is to generate traces that mimic the original traces on different statistical measures and patterns to a close degree. Conversely, we would not like the model to memorize long sequences from the original trace and place them in the generated trace. Applications trained on generated traces would benefit more from a varied trace than one that copies the original. 
However, since we deal with traces from different domains and traffic patterns, some traces may be less varied. Thus, much like with natural language generation \cite{mccoy2023much}, not all memorization is strictly bad.  
For example, a trace with only a few sequences forces the model to copy existing sequences to some degree. Ultimately, the degree of novelty of generated traces will depend on the original source trace and the user's needs. 

In this section, we will analyze the novelty of generated trace by looking at \emph{n-grams}. These are $n$-long consecutive sequences of elements. For our context, we define each element in the n-gram as a request, $(s,d)$, a pair of source and destination IDs. Since our traces generated traces are rather long, we randomly sample a set of $6000$ n-grams of each length and search for them in the original trace. 

A similar methodology was used for generative LLM to estimate the uniqueness of the generated text \cite{mccoy2023much,carlini2022quantifying} as well as in Kong \textit{et al.} \cite{kong2024high} when evaluating generated cellular network traces.

Figure \ref{fig:novelNgram} shows the proportion of novel n-gram in the generated trace from the original trace to n-gram length. We consider it a hit if an n-gram from the generated trace exists in its source trace at least once, and we have tested n-grams in the range $[1,40]$.  For \ref{fig:novelNgram}  (a) shows the results of the HPC set. In this set, we see that not all traces have the same behavior. For the Nekbone and MOCFE traces, we observe that nearly all n-grams larger than $n=5$ are novel, that is, around $99\%$. The Multigrid trace has a more gradual behavior where at $n=5$, only $85\%$ of n-grams are novel, while for $n=25$ around $98\%$ of n-grams are novel. The CNS trace is an outlier, where the proportion of novel n-grams is only $25\%$ at $n=5$ and $70\%$ for $n=40$ at the edge of the tested range; we review a possible interpretation for this later in this section. In \ref{fig:novelNgram}  (b), shows the results for the Facebook set. We see that this set has a more uniform behavior when compared to the HPC set. In each trace for n-grams larger than $n=7$, nearly all n-grams are novel, that is  $99.9\%$. 

\begin{figure}[h]
  \begin{centering}
   \begin{tabular}{c} 
  \subcaptionbox{HPC}{\includegraphics[width=0.43\textwidth]{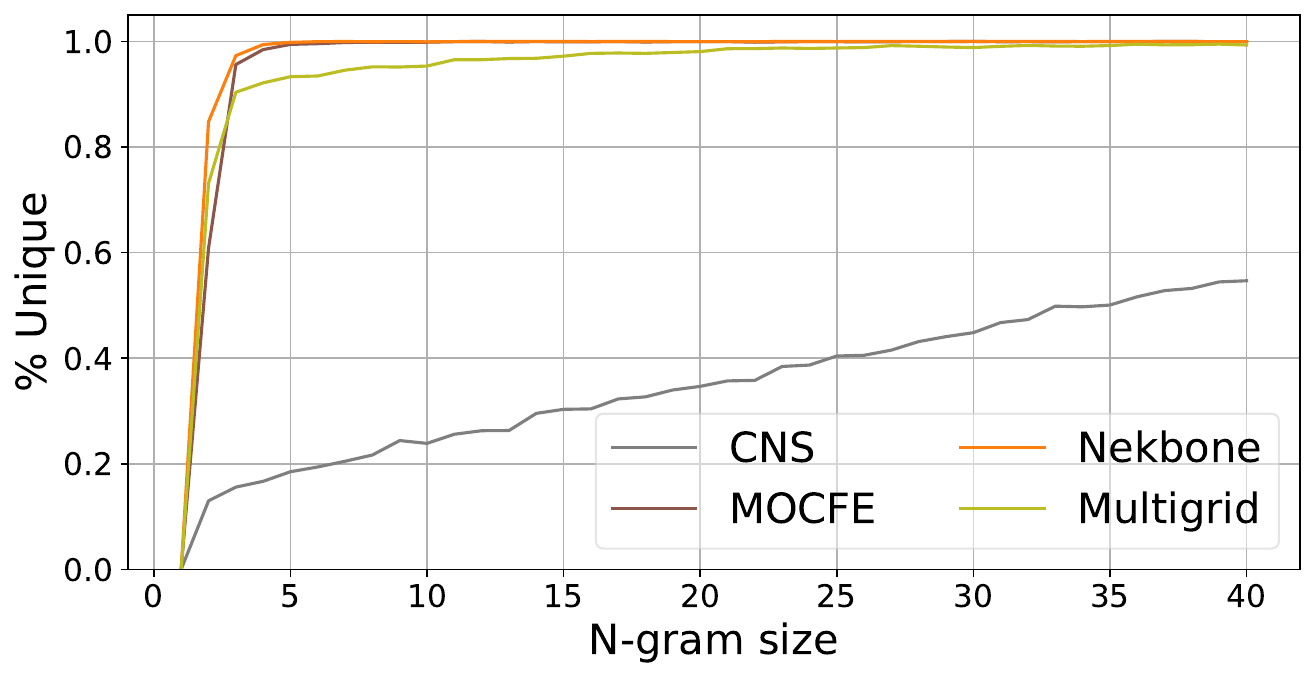}}
\end{tabular}
    \caption{ Proportion of n-grams that are unique against n-gram size between the original and generated trace in for HPC,  }
    \label{fig:uniqueNgram}
  \end{centering}
\end{figure}

Returning to the CNS trace, we suspect the reason for its different behavior is the structure of the traces themselves.
In Figure \ref{fig:uniqueNgram}, we look at the proportion of unique n-grams to n-gram length in each trace from the HPC set. Here, we consider a hit if an n-gram appears more than once. 
We see that, indeed, the proportion of unique n-grams in every trace is similar to the results of Figure \ref{fig:novelNgram}  (a). Traces other than CNS are mostly unique after $n=5$, but CNS is not. The Multigrid trace also shows a lesser degree of uniqueness when compared to the MOCFE and Nekbone traces, as before.

To conclude, other than for the CNS trace, traces generated from our \model model exhibit a high degree of novelty. We believe this pattern closely follows our result In Figure \ref{fig:uniqueNgram}, which shows a lack of novelty in the original CNS trace. Indeed, our model seems to generate novel traces that follow the 'novelty degree' of their original counterparts. 
Lastly, we also note that for every trace, there are at least a few long n-grams that have been copied from the original trace. This is likely expected to occur when using the GPT architecture and is also a known phenomenon in large language models \cite{mccoy2023much}. 

%%%%%%%%%%%%%%%%%%%%%%%%%%%%%%%%%%%%%%%%%%%%%%%%%%%%%%%%%%%%%%%%%%%%%%%%%%%%%%%%%%%%%%%%%%%%%%%%%%%%%%%%%%%%%%%%%%%%%%%%%%%%%%%%%%%%%%%%%%%%%%%%%%%%%%%%%
%%%%%%%%%%%%%%%%%%%%%%%%%%%%%%%%%%%%%%%%%%%%%%%%%%%%%%%%%%%%%%%%%%%%%%%%%%%%%%%%%%%%%%%%%%%%%%%%%%%%%%%%%%%%%%%%%%%%%%%%%%%%%%%%%%%%%%%%%%%%%%%%%%%%%%%%%%%%%%%
%%%%%%%%%%%%%%%%%%%%%%%%%%%%%%%%%%%%%%%%%%%%%%%%%%%%%%%%%%%%%%%%%%%%%%%%%%%%%%%%%%%%%%%%%%%%%%%%%%%%%%%%%%%%%%%%%%%%%%%%%%%%%%%%%%%%%%%%%%%%%%%%%%%%%%%%%%%%%%%
%%%%%%%%%%%%%%%%%%%%%%%%%%%%%%%%%%%%%%%%%%%%%%%%%%%%%%%%%%%%%%%%%%%%%%%%%%%%%%%%%%%%%%%%%%%%%%%%%%%%%%%%%%%%%%%%%%%%%%%%%%%%%%%%%%%%%%%%%%%%%%%%%%%%%%%%%%%%%%%
%%%%%%%%%%%%%%%%%%%%%%%%%%%%%%%%%%%%%%%%%%%%%%%%%%%%%%%%%%%% %%%%%%%%%%%%%%%%%%%%%%%%%%%%%%%%%%%%%%%%%%%%%%%%%%%%%%%%%%%%%%%%%%%%%%%%%%%%%%%%%%%%%%%

\section{Conclusions and Future Work}\label{sec:FutureWork}
In this paper, we proposed \model, a GPT-based machine learning model for synthetic datacenter traffic generation. 
We have also proposed several measures to judge the fidelity of the generated trace. 
Our evaluation shows \model can generate novel traces from different domains, lengths, and network sizes while maintaining fidelity. We have shown that \model can generate traces which maintain a nontrivial temporal pattern similar to the original traces. We've also briefly explored the effect of temperatures on the generated traces.    
However, our work only partially solves the problem of synthetic datacenter traffic generation.
The lack of real traces from datacenter operators continues to be a major hurdle, as no significant progress can be made without more data.% We Hope that models such \model enable companies to release their traces via the models themselves.
Evidently, by Section \ref{subsec:novelty}, nearly all sequences produced by our model are novel. We, therefore, hope that in the future, DCN operators will be able to release their network traces indirectly as a set of weights to a model, as these will hopefully not infringe on privacy laws \cite{sicker2007legal}. 
Our work takes some preliminary steps to evaluate generated traces. We acknowledge that further work will be required to improve the evaluation of new synthetic traces' fidelity by introducing other and more rigorous measurements, possibly by examining the development of the demand graph over time and testing the model's adherence to the rules of different communication protocols. This, in turn, will allow the development of better models. %rigorous

Further work can be done to improve and expand on \model.
Expanding the token set of the model to include other fields in a network trace. Other address fields (Mac address, ports, etc.) could be tokenized similarly, as the range of these addresses is limited. Increasing the number of tokens in the field vector (as discussed in Section \ref{sec:Architecture}) should make this addition natural. We believe a GPT-based architecture could be similarly capable of recreating more complex traces containing those fields.  

The time field may be harder to mimic. however, we would consider employing one-hot encoding as was done in this paper by Kong et al. \cite{kong2024high}.

To conclude, we believe that in the future, GPT-based large-scale network traffic models may be able to create most types of network traces with a large degree of detail.
 
An important step towards this goal would be training on a very large-scale dataset, which would be possible. This will enable researchers and network architects to design and develop future datacenter networks.

%%%%%%%%%%%%%%%%%%%%%%%%%%%%%%%%%%%%%%%%%%%%%%%%%%%%%%%%%%%%%%
%%%%%%%%%%%%%%%%%%%%%%%%%%%%%%%%%%%%%%%%%%%%%%%%%%%%%%%%%%%%%%
%%%%%%%%%%%%%%%%%%%%%%%%%%%%%%%%%%%%%%%%%%%%%%%%%%%%%%%%%%%%%%
%%%%%%%%%%%%%%%%%%%%%%%%%%%%%%%%%%%%%%%%%%%%%%%%%%%%%%%%%%%%%%
%%%%%%%%%%%%%%%%%%%Related work%%%%%%%%%%%%%%%%%%%%%%%%%%%%%%% 
%%%%%%%%%%%%%%%%%%%%%%%%%%%%%%%%%%%%%%%%%%%%%%%%%%%%%%%%%%%%%%
%%%%%%%%%%%%%%%%%%%%%%%%%%%%%%%%%%%%%%%%%%%%%%%%%%%%%%%%%%%%%%
%%%%%%%%%%%%%%%%%%%%%%%%%%%%%%%%%%%%%%%%%%%%%%%%%%%%%%%%%%%%%%
%%%%%%%%%%%%%%%%%%%%%%%%%%%%%%%%%%%%%%%%%%%%%%%%%%%%%%%%%%%%%%

\bibliographystyle{IEEEtran} 
\bibliography{bibi.bib}
\cleardoublepage 
%}
%\appendix \label{appendix}
\appendix

%\section*{Appendices}

\begin{figure}[!ht]
  \begin{centering}
  \begin{tabular}{c}
  \includegraphics[width=.70\linewidth]{Figs/MOCFE_orig_matrix_max_100.pdf} \\
  \small{(a) Original}\\
  \includegraphics[width=.70\linewidth]{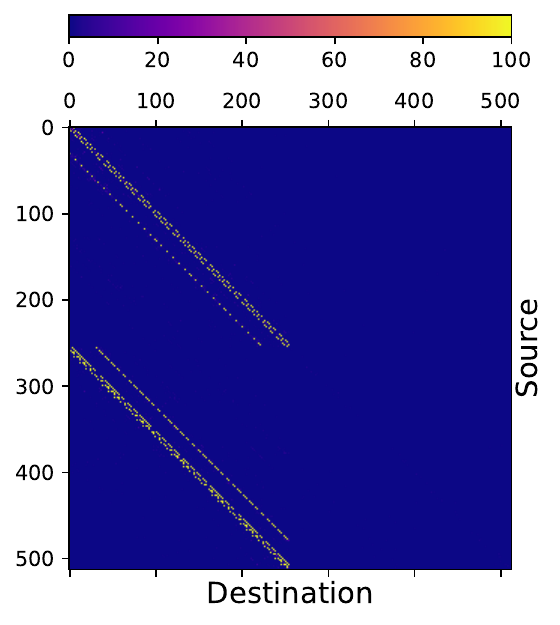} \\
    \small{(b) Generated} \\
  \includegraphics[width=.70\linewidth]{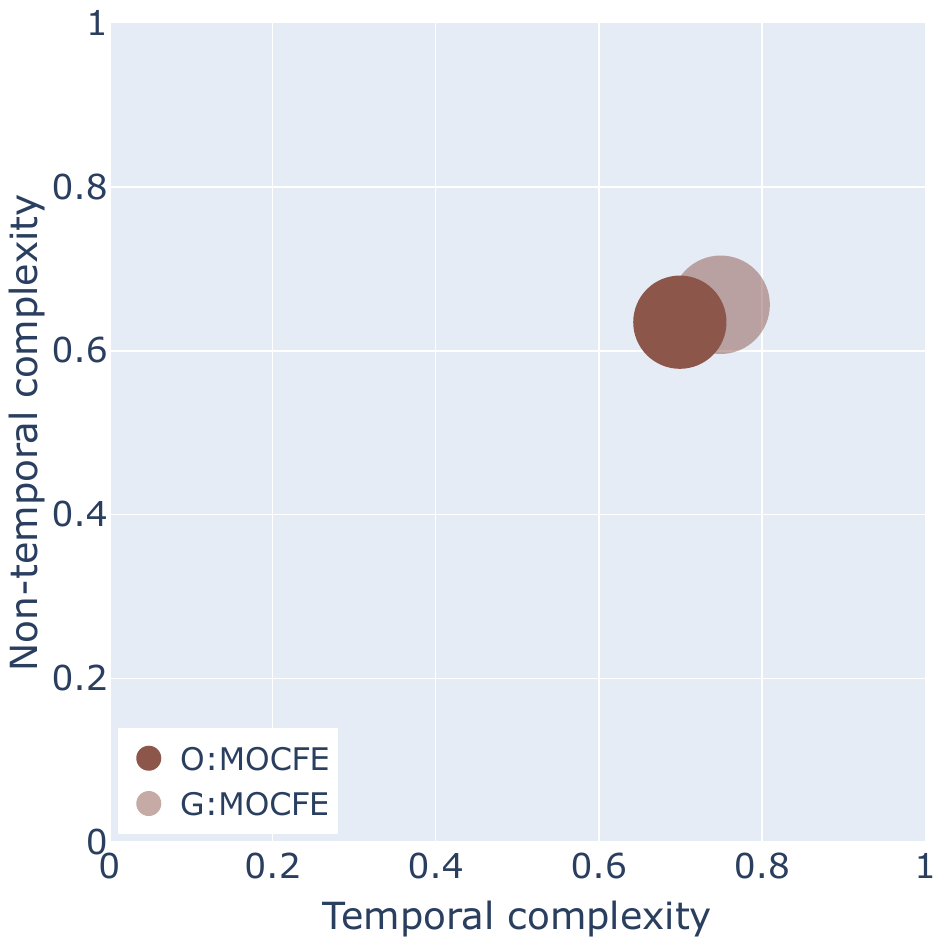} \\
   \small{(b) Complexity map}
  \end{tabular}
   \caption {Two traffic matrices of $8M$ requests. Where (a) is the original Neckbone trace\cite{doe2016characterization} The maximal element was clipped to $100$ requests to enhance contrast.  }
    \label{fig:smallTrace}
  \end{centering}
 % \vspace{-.4cm}
\end{figure}

An interesting use case for our approach is that it also lends itself to generating traces with networks of smaller scale, i.e., having fewer nodes. Throughout the rest of the paper, all generated traces have been generated to have exactly the same number of source and destination nodes as their original counterparts (as mentioned in Table \ref{tab:traces_data} ). However, generating a trace with a smaller set of nodes is simple through output masking of the logits.
A full review of the results for smaller-scale traces generated from our \model is beyond the scope of this paper. However, we provide a small test case that may hint towards further capabilities and have recreated the MOCFE traces with $512$ nodes instead of the original $1024$. 
Figure \ref{fig:smallTrace} shows the results of our test on two matrices tests in this paper, traffic matrix similarity and trace complexity.
\ref{fig:smallTrace} (b) shows the traffic matrix for the smaller-scale generated trace, and Figure \ref{fig:smallTrace} (a) shows the original traffic matrix for the MOCFE trace for reference.
At first glance, the generated traces show a similar pattern to the original. Both have main diagonals composed of smaller diagonals. It may also seem Figure \ref{fig:smallTrace} (b) is merely a zoomed-in version of the matrix in Figure \ref{fig:smallTrace} (a). However, interestingly, it seems that the two diagonals in the matrix in Figure \ref{fig:smallTrace} (b) are actually the upper and lower diagonals from the matrix in Figure \ref{fig:smallTrace} (a), and these also contain some smaller differences.
\ref{fig:smallTrace} (c) shows the complexity map for the generated and original traces. We recall that the trace complexity approach is agnostic to the scale of the network generating the trace \cite{complexity2020}; thus, we can compare the two traces on the same map. From the map, we can see that the generated trace, while smaller, can mimic the complexity profile of the original trace.

To conclude, this short test case has interestingly shown that, at least in this case, our \model model can generate a trace of a smaller scale that mimics the original trace's traffic pattern to some degree. Further research into different traces and different scales will be required, as well as the development of normalized measurements.
  
\section{Traces for Smaller Scale Networks}\label{app:smallScale}
 \end{document}